\documentclass[twocolumn]{aastex701}

\usepackage{soul}
\definecolor{green2}{rgb}{0.2,0.7,0.1}

\newcommand\N[1]{{\color{blue} \bf #1}}

\DeclareRobustCommand{\VAN}[3]{#2}
\let\VANthebibliography\thebibliography
\def\thebibliography{\DeclareRobustCommand{\VAN}[3]{##3}\VANthebibliography}
\usepackage[percent]{overpic}
\usepackage{xcolor} 
\usepackage{placeins}
\usepackage{soul}
\usepackage{booktabs}
\usepackage{xspace}
\newcommand{\xmm}{\textit{XMM-Newton}\xspace}
\newcommand{\chandra}{\textit{Chandra}\xspace}

\usepackage{threeparttable}

\begin{document}

\title{
AT~2016blu: Accretion-Powered Outbursts in a Luminous Blue Variable and Compact Object Binary
}

\author[0000-0001-8341-3940]{Mojgan Aghakhanloo}
\affiliation{Department of Astronomy, University of Virginia, 530 McCormick Road, Charlottesville, VA 22904, USA}
\affiliation{Virginia Institute of Theoretical Astronomy, University of Virginia, Charlottesville, VA 22904, USA}
\email[show]{mvy4at@virginia.edu}

\author[0000-0003-3638-8943]{Núria Torres-Albà}
\affiliation{Department of Astronomy, University of Virginia, 530 McCormick Road, Charlottesville, VA 22904, USA}
\email[show]{hzr5bt@virginia.edu}

\author[0000-0001-5510-2424]{Nathan Smith}
\affiliation{Steward Observatory, University of Arizona, Tucson, 85721, AZ, USA}
\email{}

\author[0009-0004-7268-7283]{Raphael Baer-Way}
\affiliation{Department of Astronomy, University of Virginia, 530 McCormick Road, Charlottesville, VA 22904, USA}
\email{}

\author[0000-0002-1856-9225]{Shazrene Mohamed}
\affiliation{Department of Astronomy, University of Virginia, 530 McCormick Road, Charlottesville, VA 22904, USA}
\affiliation{Virginia Institute of Theoretical Astronomy, University of Virginia, Charlottesville, VA 22904, USA}
\affiliation{South African Astronomical Observatory, P.O Box 9, Observatory, 7935, Cape Town, South Africa}
\affiliation{Department of Astronomy, University of Cape Town, Private Bag X3, Rondebosch, 7701, Cape Town, South Africa}
\affiliation{NITheCS National Institute for Theoretical and Computational Sciences, South Africa}
\email{}

\author[0000-0003-3460-0103]{Alexei V. Filippenko}
\affiliation{Department of Astronomy, University of California, Berkeley, 94720-3411, CA, USA}
\email{}

\author[0000-0002-2636-6508]{WeiKang Zheng}
\affiliation{Department of Astronomy, University of California, Berkeley, 94720-3411, CA, USA}
\email{}

\author[0000-0001-5955-2502]{Thomas G. Brink }
\affiliation{Department of Astronomy, University of California, Berkeley, 94720-3411, CA, USA}
\email{}


\author[0000-0003-4253-656X]{D. Andrew Howell}
\affiliation{Las Cumbres Observatory, 6740 Cortona Drive, Goleta, 93117-5575, CA, USA}
\affiliation{Department of Physics, University of California, Santa Barbara, 93106-9350, CA, USA}
\email{}

\author[0000-0002-9454-1742,gname=Brian,sname=Hsu]{Brian~Hsu}
\affiliation{Steward Observatory, University of Arizona, Tucson, 85721, AZ, USA}
\email{bhsu@arizona.edu}

\author[0000-0002-1895-6639]{Moira Andrews}
\affiliation{Las Cumbres Observatory, 6740 Cortona Drive, Goleta, 93117-5575, CA, USA}
\affiliation{Department of Physics, University of California, Santa Barbara, 93106-9350, CA, USA}
\email{}

\author[0000-0003-4914-5625]{Joseph R. Farah}
\affiliation{Las Cumbres Observatory, 6740 Cortona Drive, Goleta, 93117-5575, CA, USA}
\affiliation{Department of Physics, University of California, Santa Barbara, 93106-9350, CA, USA}
\email{}

\author[0000-0001-5807-7893]{Curtis McCully}
\affiliation{Las Cumbres Observatory, 6740 Cortona Drive, Goleta, 93117-5575, CA, USA}
\email{}

\author[]{Kathryn Wynn}
\affiliation{Las Cumbres Observatory, 6740 Cortona Drive, Goleta, 93117-5575, CA, USA}
\affiliation{Department of Physics, University of California, Santa Barbara, 93106-9350, CA, USA}
\email{}

\author[0000-0002-8770-6764]{R\'eka K\"onyves-T\'oth}
\affiliation{Konkoly Observatory, HUN-REN Research Center for Astronomy and Earth Sciences, Konkoly
Th. M. út 15-17., Budapest, 1121 Hungary; MTA Center of Excellence}
\email{konyvestoth.reka@csfk.org}

\author[orcid=0000-0003-4175-4960]{Conor~L.~Ransome}
\affiliation{Steward Observatory, University of Arizona, Tucson, 85721, AZ, USA}
\email{cransome@arizona.edu} 

\author[]{Enrique Rathmann}
\affiliation{Observatorio Tros-Alt, Monserrat, Valencia, Spain}
\email{cara-norte@hotmail.com}

\author{Tam\'as Tordai}
\affiliation{Polaris Observatory, Hungarian Astronomical Association, Laborc utca 2/c, 1037 Budapest, Hungary}
\email{}

\author[]{Sjoerd Dufoer}
\affiliation{Vereniging Voor Sterrenkunde (VVS), Zeeweg 96, 8200 Brugge, Belgium}
\email{}

\author[]{Charles Cynamon}
\affiliation{Supra Solem Observatory, Bradley, CA}
\email{}

\author[0000-0003-0125-8700]{Franz-Josef Hambsch}
\affiliation{Groupe Européen d’Observations Stellaires (GEOS), 23 Parc de Levesville, 28300 Bailleau l’Evêque, France}
\affiliation{Bundesdeutsche Arbeitsgemeinschaft für Veränderliche Sterne (BAV), Munsterdamm 90, 12169 Berlin, Germany}
\affiliation{Vereniging Voor Sterrenkunde (VVS), Zeeweg 96, 8200 Brugge, Belgium}
\email{}

\author[0009-0002-4296-2200]{Thomas Rutherford}
\affiliation{American Association of Variable Star Observers, citizen amateur (AAVSO), 185 Alewife Brook Parkway, Suite 410, Cambridge, MA 02138, USA}
\affiliation{Department of Physics and Astronomy, East Tennessee State University, 1276 Gilbreath Drive, Johnson City, TN 37614, USA}
\affiliation{Department of Natural Sciences and Mathematics, King University, 1350 King College Road, Bristol, TN 37620}
\email{}

\author[0009-0005-0424-1842]{Simone Leonini}
\affiliation{Montarrenti Observatory, Strada Provinciale 73 bis, 53018, Sovicille, Italy} 
\affiliation{American Association of Variable Star Observers, citizen amateur (AAVSO), 185 Alewife Brook Parkway, Suite 410, Cambridge, MA 02138, USA}
\affiliation{UAI, Unione Astrofili Italiani, Parco Astronomico ``Livio Gratton”, Via Lazio 14 - localit`a Vivaro, 00040, Rocca di Papa, RM, Italy}
\email{}

\author[]{Franky Dubois}
\affiliation{Vereniging Voor Sterrenkunde (VVS), Zeeweg 96, 8200 Brugge, Belgium}
\affiliation{Amateur observatory ``Adonis" Poelkapellestraat 57, 8920 Langemark, Belgium}
\email{}

\author[0000-0003-3833-2695]{Ivo Peretto}
\affiliation{MarSEC, Marana Space Explorer Center, Pasquali Road, 36073, Crespadoro, VI, Italy}
\affiliation{American Association of Variable Star Observers, citizen amateur (AAVSO), 185 Alewife Brook Parkway, Suite 410, Cambridge, MA 02138, USA}
\affiliation{UAI, Unione Astrofili Italiani, Parco Astronomico ”Livio Gratton”, Via Lazio 14 - localit`a Vivaro, 00040, Rocca di Papa, RM, Italy}
\email{}

\author[]{Donald F. Collins}
\affiliation{College View Observatory, 138 College View Dr., Swannanoa, NC 28778, USA}
\email{}

\author[0009-0008-2210-8924]{Lorenzo Betti}
\affiliation{Osservatorio Polifunzionale del Chianti (OPC) - University of Florence}
\affiliation{Strada Provinciale 101 Castellina in Chianti, 9.25 km, 50028 Barberino Tavarnelle,  Florence, Italy}
\email{}

\author[0009-0002-7967-6830]{Lara Fossi}
\affiliation{Osservatorio Polifunzionale del Chianti (OPC) - University of Florence}
\affiliation{Strada Provinciale 101 Castellina in Chianti, 9.25 km, 50028 Barberino Tavarnelle,  Florence, Italy}
\email{}

\author[0009-0005-4783-0049]{Irene Parenti}
\affiliation{Osservatorio Polifunzionale del Chianti (OPC) - University of Florence}
\affiliation{Strada Provinciale 101 Castellina in Chianti, 9.25 km, 50028 Barberino Tavarnelle,  Florence, Italy}
\email{}

\author[]{Vincenzo Pettina}
\affiliation{Osservatorio Polifunzionale del Chianti (OPC) - University of Florence}
\affiliation{Strada Provinciale 101 Castellina in Chianti, 9.25 km, 50028 Barberino Tavarnelle,  Florence, Italy}
\email{}

\author[]{Giulio Giordano Chiti}
\affiliation{Osservatorio Polifunzionale del Chianti (OPC) - University of Florence}
\affiliation{Strada Provinciale 101 Castellina in Chianti, 9.25 km, 50028 Barberino Tavarnelle,  Florence, Italy}
\email{}

\author[]{Maria Hofheinz}
\affiliation{Osservatorio Polifunzionale del Chianti (OPC) - University of Florence}
\affiliation{Strada Provinciale 101 Castellina in Chianti, 9.25 km, 50028 Barberino Tavarnelle,  Florence, Italy}
\email{}

\author[]{André Steenkamp}
\affiliation{Southwater Observatory, Southwater, Horsham, United Kingdom}
\email{}

\author[]{Sauro Gaudenzi}
\affiliation{American Association of Variable Star Observers, citizen amateur (AAVSO), 185 Alewife Brook Parkway, Suite 410, Cambridge, MA 02138, USA}
\email{}

\author[0009-0002-1484-7029]{Nikola Antonov}
\affiliation{Institute for Advanced Physical Studies, 111 Tsarigradsko Shose Blvd., Sofia, Bulgaria}
\affiliation{Department of Astronomy, Sofia University, 5 James Bourchier Blvd., Sofia, Bulgaria}
\email{}

\author{Giuseppe Pappaa}
\affiliation{American Association of Variable Star Observers, citizen amateur (AAVSO), 185 Alewife Brook Parkway, Suite 410, Cambridge, MA 02138, USA}
\email{giuseppe.pappa@alice.it}

\email{}



\begin{abstract}
We present the first X-ray detection of the supernova (SN) impostor AT~2016blu in NGC~4559. AT~2016blu exhibited 27 detected quasiperiodic ($\sim$113 day) outbursts between 2012 and June 2026, presumably triggered by periastron passages in an eccentric binary system in which the primary is a $\gtrsim 33~M_{\odot}$ luminous blue variable (LBV). AT~2016blu was serendipitously observed by \textit{Chandra} in 2001–2002, prior to its first documented outburst in 2012, but the stacked archival data show no X-ray detection. We monitored the source's visual-wavelength variability around its predicted outburst time and triggered a \textit{Chandra} Target of Opportunity program in March 2026, obtaining observations with detections over five closely spaced epochs. These data indicate an X-ray luminosity of $\log L_{\rm X} \approx 38.64 \pm 0.11$~erg~s$^{-1}$ and an accretion rate of $\dot{M} \gtrsim 8\times10^{-8}~M_{\odot}~\mathrm{yr^{-1}}$, consistent with the presence of a compact companion. We therefore conclude that AT~2016blu is the first known case of an LBV SN impostor whose outbursts are driven by intermittent accretion onto a compact object. Given that the system consists of a massive star and a compact companion, AT~2016blu is a high-mass X-ray binary, similar to SN~2010da, although the donor star in SN~2010da is a red supergiant.

\end{abstract}



\section{Introduction} \label{sec:intro}

Supernova (SN) impostors are transient events associated with massive stars that are often initially discovered in SN surveys \citep{V00,S11}. However, because their outbursts are less energetic and less luminous than true supernovae (SNe), and because the star is expected to survive, they are subsequently classified as SN impostors. These events are usually linked to the eruptions of luminous blue variables (LBVs), which are evolved, unstable, massive stars characterized by episodic mass loss and photometric variability \citep{smith26}. The LBV phenomenon encompasses a range of variability, including giant eruptions, such as that of $\eta$ Carinae \citep{smith18}, and S~Doradus-type variability  \citep{H99,V01,V05}. Giant eruptions correspond to extreme mass-loss episodes of at least ${\sim}0.01$--$1~\mathrm{M_{\odot}~yr^{-1}}$, while S~Doradus-type variability is associated with 
more moderate mass-loss rates of 
${\sim}10^{-5}$--$10^{-4}~\mathrm{M_{\odot}~yr^{-1}}$ 
\citep{D96, V02, G09, S14}. This mass loss may play a key role in the formation of Wolf–Rayet (W-R) stars and in shaping the dense circumstellar material (CSM) observed around interacting SNe.

In single-star evolutionary models, LBVs are considered a transitional phase preceding the W-R stage and are not expected to explode directly as SNe \citep{C98}. However, growing observational evidence challenges this picture \citep{GL09,so06,smith07,S10,M13,B22,J22,S22}. A prominent example is SN~2009ip, which had an LBV-like progenitor and later exploded as an SN IIn, in tension with theoretical expectations \citep{S10,M13}. This discrepancy with a single-star scenario should be expected, as an increasing number of studies indicate that a majority of massive stars are affected by binary interaction \citep{S12, M17}. In this context, LBVs may arise from binary evolution channels, as suggested by their observed spatial distributions, kinematics, and local environments \citep{S15, smith16, smith19, A17, A22, A26}. In particular, mass accretion or stellar mergers can result in a star with a helium core that is small relative to its total mass and a much larger hydrogen-rich envelope. Although such systems often have longer total lifetimes than equivalent-mass single stars, this structure keeps them in the blue part of the HR diagram as blue stragglers. They can therefore explode as hydrogen-rich, LBV-like progenitors without ever passing through a W-R phase \citep{J14,S24}.

Recently, two SN impostors, AT~2016blu and SN~2000ch, have shown similarities to both the precursor activity of SN~2009ip and outbursts of $\eta$ Carinae before its giant eruption.  AT~2016blu and SN~2000ch exhibit repeated nonterminal outbursts over the past one to three decades, with quasiperiodic behavior that may be driven by periastron interactions in eccentric binary systems, where the primary star is an LBV \citep{A2300ch,A2316blu}. However, the exact nature of these periastron interactions and the properties of the companion star remain uncertain. 

In this paper, we focus on AT~2016blu in NGC~4559 at a distance of $8.91 \pm 0.3$~Mpc \citep{Mc17} with a redshift of $z=0.00261$.  AT~2016blu experienced its first recorded outburst in 2012  \citep{K12}. In \citet{A2316blu}, using data from the 0.76\,m Katzman Automatic Imaging Telescope \citep[KAIT;][]{F01} at Lick Observatory obtained during monitoring of the NGC~4559 galaxy, we found that the system was stable prior to its first outburst in 2012. AT~2016blu, while in a stable state (in 2001 and 2005), was observed with the Hubble Space Telescope (HST; \citealt{V12}); these observations implied a mass and luminosity of $M \gtrsim 33$~M${_\odot}$ and $L \gtrsim 10^{5.7}$~L${_\odot}$. In \citet{A2316blu}, we concluded that AT~2016blu is a massive blue straggler, as its mass is inconsistent with that of the surrounding massive stars. Using data from multiple telescopes, we identified at least 21 outbursts, although additional events were likely missed during the times that were difficult to observe the target owing to periods when the source was in solar conjunction. 

In \citet{A25},  we also analyzed AT~2016blu’s spectra and found strong H$\alpha$ and He~{\sc i} emission, consistent with a hot LBV and closely resembling the precursor spectrum of SN~2009ip. AT~2016blu’s spectra also exhibit several notable features, including high-velocity P~Cygni absorption features and multicomponent absorption structures, none of which show a clear correlation with orbital phase or magnitude. This suggests that processes beyond simple binary interaction, such as interaction with a companion’s jet, accretion disk, or a clumpy CSM, may play a role. The optical light curve further exhibits multiple narrow peaks of comparable magnitude within individual eruptive episodes, which may also be linked to such processes, while the sequence of peaks is quasiperiodic, likely due to the binary nature of the system.

While the light curve and spectral analyses of AT~2016blu establish its binary nature, they cannot determine the nature of its companion and the underlying periastron interactions (e.g., wind–wind collisions, accretion onto a compact object). 
The most direct way to probe the interactions is through X-ray observations. In a wind-wind collision scenario, the winds of two massive stars collide, producing shocks that heat gas and radiates thermally at moderate X-ray luminosities. In an accretion scenario, by contrast, material captured from the primary's wind releases gravitational potential energy as it falls onto a compact companion, producing hard X-ray emission that can be orders of magnitude more luminous \citep{D73, Naze12}.  These distinct X-ray signatures thus reveal the nature of periastron encounters and the companion. 

This paper is organized as follows. In Section~\ref{sec:obs}, we present the X-ray and optical observations and data reduction.  The results are described in Section~\ref{sec:result}. Section~\ref{sec:discussion} examines the association between the X-ray detection and AT~2016blu, and places it in context with other SN impostors. We conclude with a summary in Section~\ref{sec:summary}.

\section{Observations}\label{sec:obs}

\subsection{X-ray Observations and Data Reduction}\label{sec:x-ray_obs}

\begin{figure*}
    \centering
    \includegraphics[width=0.95\textwidth, trim=0 0 0 0, clip]{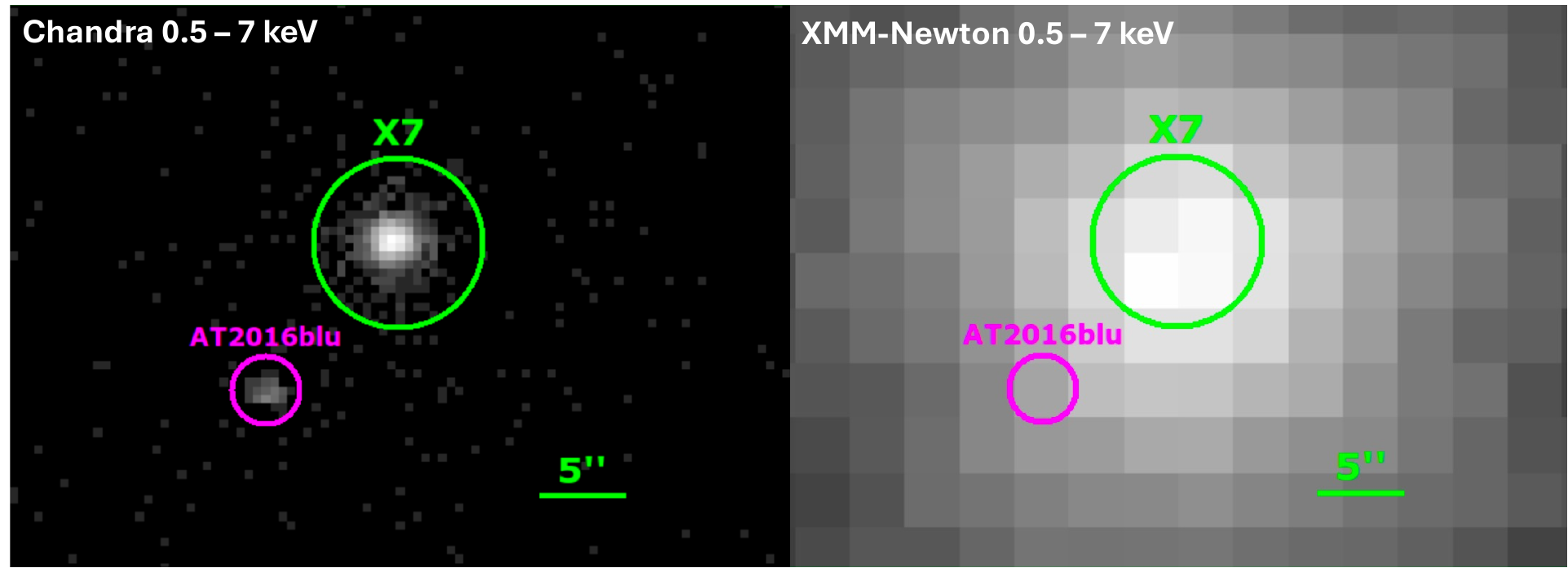}
    \caption{{\it Left:} Stacked \chandra image of all five ToO observations taken during the March 2026 optical outburst of AT~2016blu, in a span of 4~days, as listed in Table~\ref{tab:chandra_obs}. The magenta circle shows a $2''$ radius region around the optical TNS position of AT~2016blu, while the green circle corresponds to a $5''$ radius region highlighting the position of the nearby ULX NGC~4559~X7. {\it Right:} \xmm image of the same sky region. \xmm cannot resolve the two sources, and the image is clearly dominated by ULX NGC~4559~X7, much brighter than AT~2016blu.}
    \label{fig:chandra_and_xmm}
\end{figure*}

\begin{figure*}
    \centering
    \includegraphics[width=0.495\textwidth, trim=0 0 0 0, clip]{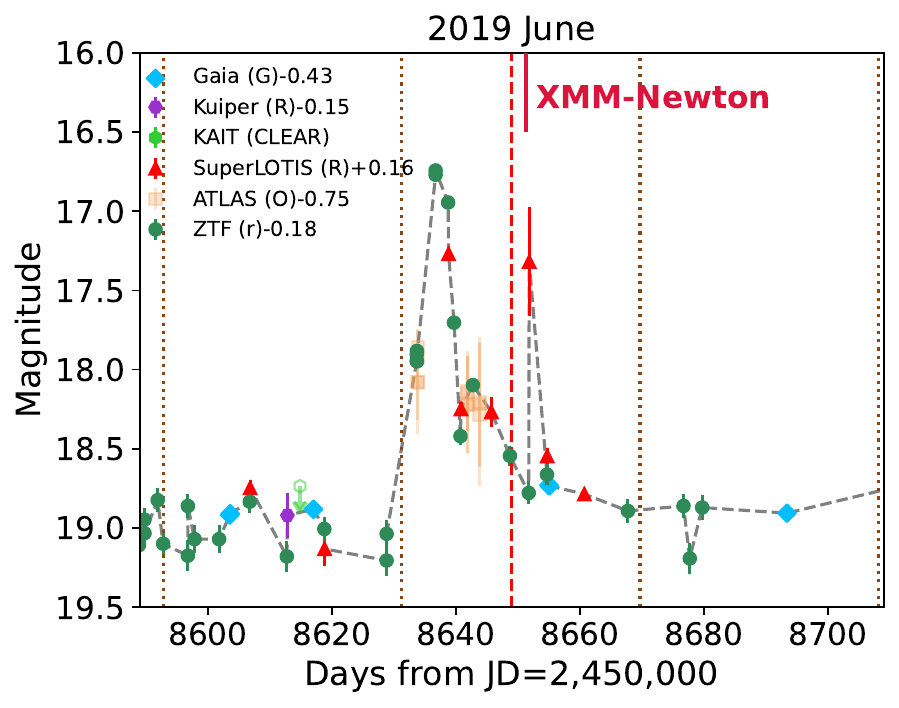}
    \hfill
    \includegraphics[width=0.495\textwidth,trim=0 0 0 0, clip]{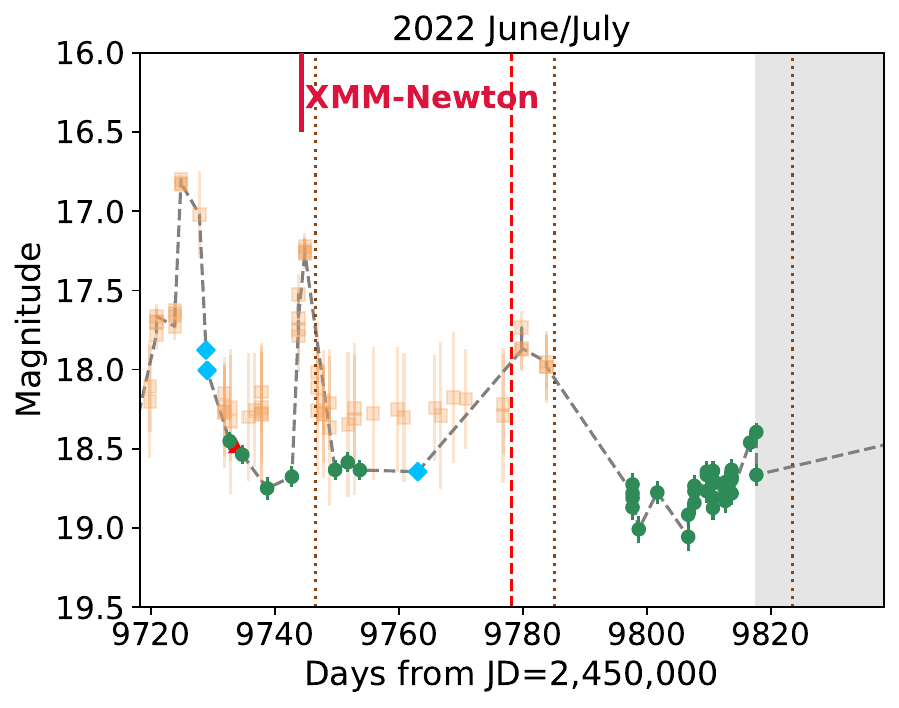}
    \caption{AT~2016blu was observed by \xmm four times, three of which occurred after its first outburst in 2012. The 2019 \xmm observation occurs just before the documented optical peak (this peak is based on a single SuperLOTIS data point with a large uncertainty and should therefore be interpreted with caution), while the two \xmm observations in 2022  correspond to AT~2016blu rising toward its optical peak, as documented by \citet{A2316blu, A25}. However, as shown in  Figure~\ref{fig:chandra_and_xmm}, owing to the low spatial resolution of \xmm, we cannot spatially resolve AT~2016blu from the nearby ULX NGC~4559~X7. Predicted outburst times from the $\sim113$ and $\sim38.5$ day periods are shown by the red dashed and brown dotted lines, respectively; gray shaded regions mark periods of limited observability due to the Sun \citep{A2316blu}. Following  \citet{A2316blu, A25}, offsets were applied as constant-flux subtractions; the quoted values correspond to the magnitude shifts at a baseline of 18--18.5\,mag.  The ZTF exception is the $-0.18$\,mag transformation to Johnson-Bessell $R$. The same plotting conventions are used throughout the paper.}
    \label{fig:xmm}
\end{figure*}

AT~2016blu was serendipitously observed by \chandra on three occasions (ObsIDs 2026, 2027, 2686; total exposure $\sim$ 23~ks; see Table \ref{tab:chandra_obs}) between 2001 and 2002. All of these data were taken when the system was stable; they predate the first outburst of AT~2016blu in 2012 \citep{A2316blu}. The archival \chandra observations were targeting a nearby ultraluminous X-ray source (ULX), NGC~4559~X7 \cite[e.g.,][]{Miller2004,R04,S05}.
AT~2016blu was also serendipitously observed by \xmm four times (with a total exposure of $\sim$250~ks) between 2003 and 2022. These observations also targeted the nearby ULX NGC~4559~X7. Owing to the angular resolution of \xmm, the observations cannot spatially resolve NGC~4559~X7 and AT~2016blu; see Figure~\ref{fig:chandra_and_xmm}. The post-2012 observations are close to the observed outbursts of AT~2016blu; see Figure~\ref{fig:xmm} (but we note that the optical peak near the 2019 \xmm\ observation is based on a single SuperLOTIS data point with a large uncertainty.) However, owing to the high brightness and variability of  NGC~4559~X7  \citep{Pintore2025}, it is not possible to ascertain whether AT~2016blu was emitting X-rays at a detectable level in any of the \xmm observations. 

AT~2016blu was observed with our \chandra Target of Opportunity (ToO) program in March 2026 for a total exposure time of $\sim$ 100~ks (PI M. Aghakhanloo). The observations were split into shorter segments to manage spacecraft and instrument temperatures; see Table~\ref{tab:chandra_obs} for details. The \chandra observations were obtained with ACIS-S and were triggered immediately after AT~2016blu reached a peak $r$-band magnitude of $16.6$ (see Section~\ref{sec:optical_obs} for details of the optical photometry). 

\begin{table}[h!]
    \centering
    \resizebox{\columnwidth}{!}{%
    \begin{threeparttable}
    \caption{\chandra X-ray Observations of AT~2016blu.}
    \label{tab:chandra_obs}
    \setlength{\tabcolsep}{4pt}
    \begin{tabular}{cccc}
        \toprule
        \textbf{ObsID} & \textbf{Time} & \textbf{Exp. Time\tnote{a}} & \textbf{Net Counts} \\
        & (UTC) & (ks) & \\   
        \midrule
        \multicolumn{4}{c}{Archival Observations} \\
        \midrule
        2026 & 2001-01-14 & 9.7 & $-$\\
        2027 & 2001-06-04 & 11.1 & $-$\\
        2686 & 2002-03-14 & 3.0 & $-$\\
        \midrule
        \multicolumn{4}{c}{ToO Campaign} \\
        \midrule
        31041 & 2026-03-24 & 23.8 & 15\\
        32254 & 2026-03-25 & 16.6 & 10\\
        32255 & 2026-03-26 & 24.2 & 37\\
        32256 & 2026-03-26 & 16.1 & 7\\
        32257 & 2026-03-27 & 16.1 & 7\\
        \bottomrule
    \end{tabular}
    \begin{tablenotes}
        \footnotesize
        \item[a] Exposure times correspond to the on-target time (uncorrected for dead time). The analysis, however, uses the effective exposure times after all corrections; see Figure \ref{fig:chandra_all_obs}.
    \end{tablenotes}
    \end{threeparttable}}
\end{table}

Each observation is reduced individually, following the standard \chandra data-analysis procedure, as described by the CIAO analysis guides\footnote{https://cxc.cfa.harvard.edu/ciao/} \citep{F26}. We use CIAO version 4.17 to carry out the analysis. We extract spectra for all observations in our ToO campaign, as shown in Table~\ref{tab:chandra_obs}. The source spectra are extracted using a circular region with a radius of $2''$, as shown in Figure~\ref{fig:chandra_and_xmm}. Owing to the proximity of NGC~4559~X7 ($\sim 10''$), we do not use an annular region to extract the background, but choose a circular region of $12''$ radius. 

AT~2016blu is detected in each individual \chandra observation (see Table~\ref{tab:chandra_obs} for the net counts). Figure~\ref{fig:chandra_and_xmm} shows a stacking of all 5 observations, totaling 95.4~ks of effective exposure time. The magenta circle marks the TNS (Transient Name Server) position of AT~2016blu. We adopt this coordinate rather than the SIMBAD position, as it shows better alignment with the \chandra\ emission. Note that some of the optical data (e.g., ATLAS) described in the next subsection are based on the SIMBAD coordinates, which are offset from the TNS position by $\sim1''$. The offset between the two positions is comparable to the angular resolution of these optical facilities and therefore does not affect our analysis.

For each observation, we fit one spectrum using a \texttt{tbabs*powerlaw} model. \texttt{tbabs} accounts for the obscuration in the direction of NGC~4559, originating in our own Galaxy, measured to be $N_{\rm H,gal}$=$1.43\times10^{20}$~cm$^{-2}$ \citep{HI4PI2016}. The value of this obscuration is fixed and not varied within the fits. 

Fitting was carried out using the Bayesian X-ray Analysis software package \citep[BXA;][]{Buchner2014,Buchner2016bxa}, which implements the nested sampling package \texttt{ULTRANEST} \citep{Buchner2016ultranest,Buchner2021}. The model has only two free parameters: the power-law index (or photon index, $\Gamma$) and normalization ($norm$). We assigned a uniform prior to $\Gamma$ and a log-uniform prior to the normalization, in the ranges of $\Gamma\in\left[-5,10\right]$ and $\log \left(norm\right)\in\left[-11,-2\right]$, for a full exploration of the parameter range.

After confirming the source does not exhibit strong variability during the 4 days it was observed (see Section~\ref{sec:x-ray_results}), we stack the spectra of all individual observations using the CIAO task \texttt{combine\_spectra}. Given the higher signal-to-noise ratio, we refit the spectra using the same setup as described for individual observations, to obtain tighter constraints on the luminosity and photon index. We also use a second model, an absorbed power law, or \texttt{ZTbabs*powerlaw}. The additional parameter, column density ($N_{\rm H}$), accounts for any potential obscuration of the X-rays along our line of sight, originating close to the source. We use a log-uniform prior in the range of $\log N_{\rm H}\in\left[-2,2\right]$, where $N_{\rm H}$ is in units of $10^{22}$~cm$^{-2}$.

\subsection{Optical Photometry}\label{sec:optical_obs}
AT~2016blu was monitored daily with Lick/KAIT in five filters ($B$, $V$, $R$, $I$, and clear). All KAIT images were reduced using a custom pipeline\footnote{https://github.com/benstahl92/LOSSPhotPypeline} detailed by \citet[][]{Stahl2019}. We applied an image-subtraction procedure using template images obtained during AT~2016blu's quiescent state.
Point-spread-function (PSF) photometry was obtained using {\tt DAOPHOT} \citep[][]{Stetson1987} from the {\tt IDL} Astronomy Users Library\footnote{http://idlastro.gsfc.nasa.gov/}. Apparent magnitudes were all measured in the KAIT4 natural system \citep[see][]{Stahl2019}.
Several nearby stars were chosen from the
Pan-STARRS1\footnote{http://archive.stsci.edu/panstarrs/search.php} catalog for calibration;
their magnitudes were first transformed into the  \citet{landolt92} system using the empirical prescription presented by Eq.~6 of \citet{Tonry2012}, and then transformed to the KAIT4 natural system.

AT~2016blu was also observed daily as part of the Las Cumbres Observatory Global Supernova Project  \citep[GSP;][]{B13}, using 1~m telescopes across the LCO global network in the $r$ band. The LCO photometric data were reduced using lcogtsnpipe
\citep{Valenti_LCOpipe} which calculates PSF magnitudes (following standard image calibration) after finding zero-points and color terms
\citep{Steson_daophot}.

In addition, we initiated an observing campaign under AAVSO Alert Notice 847 in 2024, inviting AAVSO observers to closely monitor AT~2016blu for any subsequent outbursts. 
AAVSO observers also provided daily monitoring of AT~2016blu during the \chandra ToO observations. In total, more than 30 observers contributed data over the course of the campaign until June 2026. The AAVSO data are not image-subtracted and are reduced with photometry software such as Tycho Tracker and IRAF \citep{T86,T93,P20}. We retain AAVSO measurements with uncertainties $<0.2$ mag, and bin the data to one point per night, adopting the median magnitude, with the uncertainty taken as the scatter among contributing observers or their median quoted error, whichever is larger.
The full dataset is publicly available through the AAVSO database at \url{https://app.aavso.org/webobs/results/?star=at+2016blu&num_results=25&obs_types=all}.


AT~2016blu was also observed by the {\it Gaia} space telescope \citep{G16} (but with photometry available until January 2024), the Zwicky Transient Facility \citep[ZTF;][]{B19}, and ATLAS \citep[Asteroid Terrestrial-impact Last Alert System;][]{T18}. We retrieved ZTF DR24 light curves, which provide PSF-fit photometry measured on individual calibrated science exposures rather than forced difference-image photometry. However, the ZTF  data contemporaneous with the \chandra observations were not publicly available at the time of the \chandra observations. 

The photometric datasets described above are obtained using different reduction approaches. For example, the KAIT data are image-subtracted, while the AAVSO, LCO, and ZTF data are not. Since we use these data to characterize the evolution and variability of AT~2016blu rather than to measure absolute fluxes, these differences do not affect our analysis. Further details on photometry and data reduction is provided by \cite{A2316blu}.

\begin{figure*}
    \centering
    \includegraphics[width=0.49\linewidth, trim=0 0 0 0, clip]{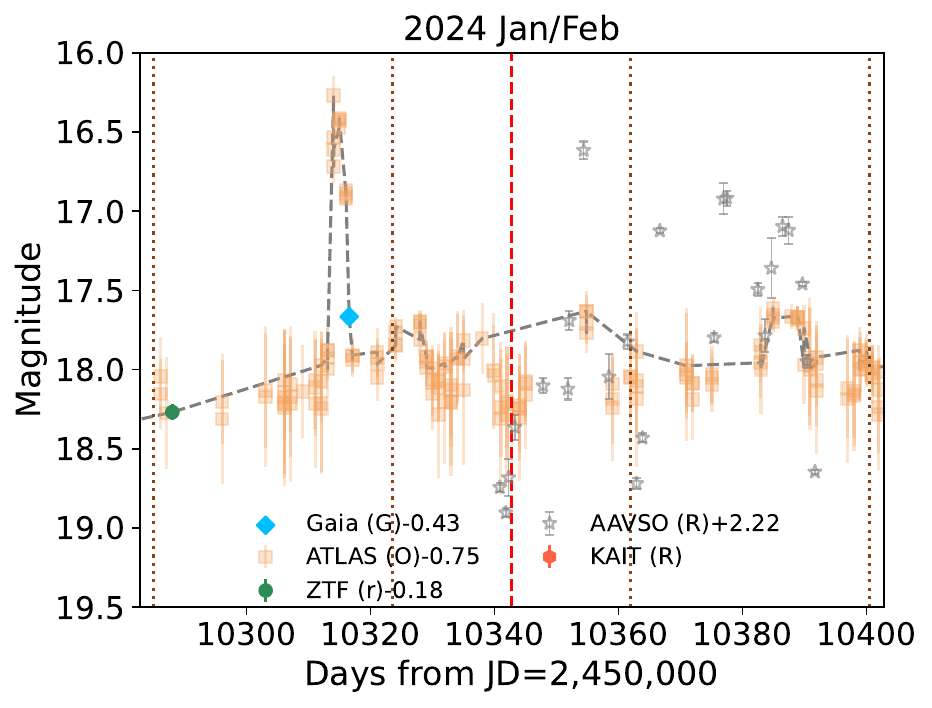}
    \hfill
    \includegraphics[width=0.49\linewidth, trim=0 0 0 0, clip]{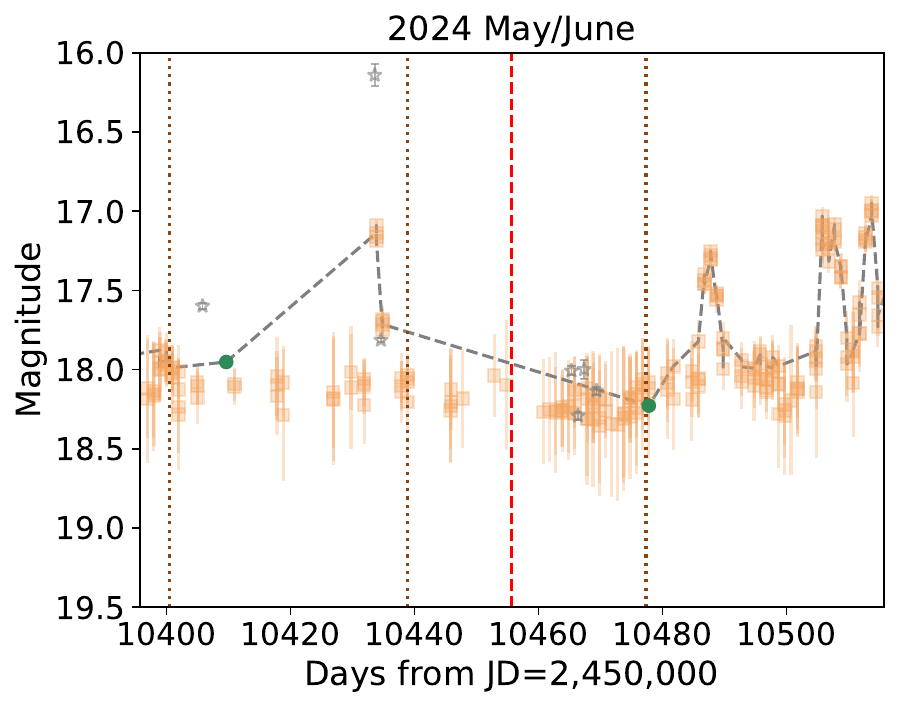}
    
    \vspace{0.5em}
    
    \includegraphics[width=0.49\linewidth, trim=0 0 0 0, clip]{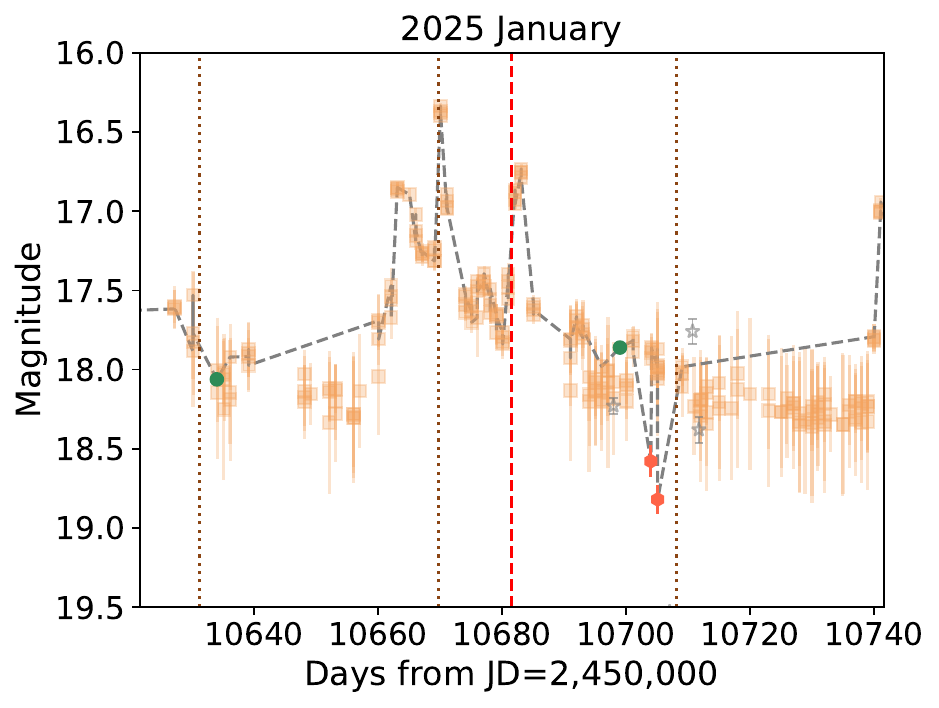}
    \hfill
    \includegraphics[width=0.49\linewidth, trim=0 0 0 0, clip]{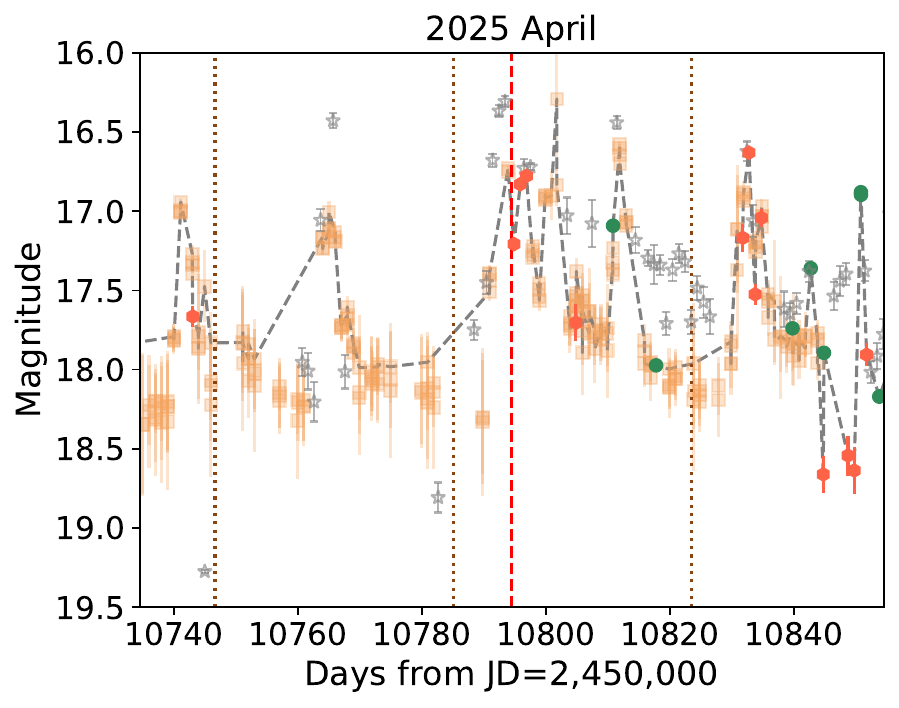}
    
    \vspace{0.5em}
    
    \includegraphics[width=0.49\linewidth, trim=0 0 0 0, clip]{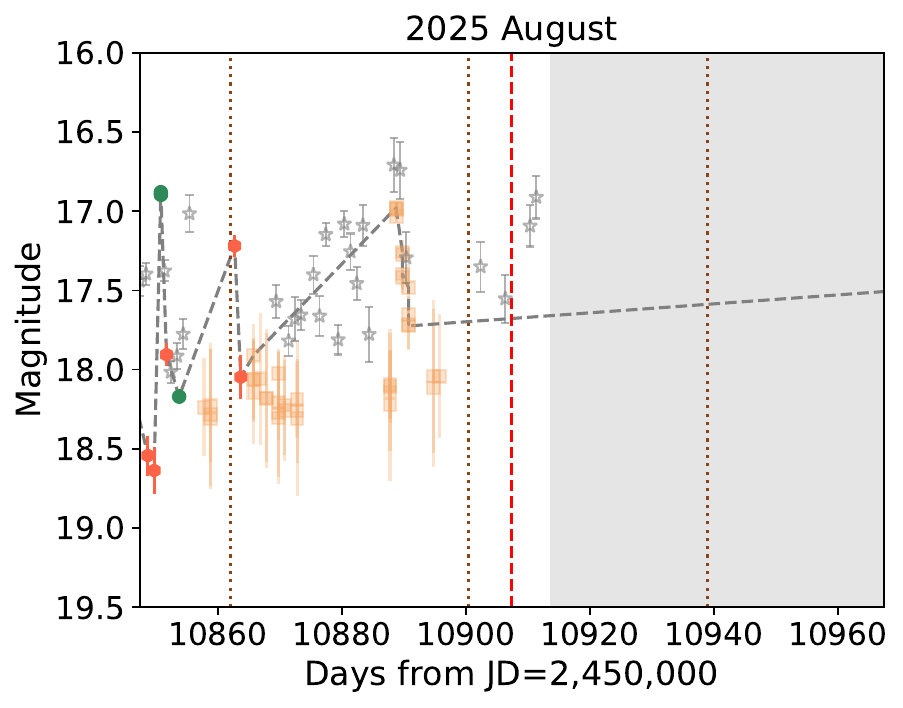}
    
    \caption{AT~2016blu has undergone five additional outbursts since its last major event in May/June 2023 \citep{A25}. Nearly all of these outbursts were monitored by AAVSO amateur astronomers (Alert Notice 847). The red dashed line indicates the expected time of the outburst
    derived from the most likely period of $\sim113$~days, tied to the January 2012 outburst. The brown dotted line marks the expected time of the outburst
    derived based on the secondary peak in the period analysis at $\sim38.5$~days \citep{A2316blu}. The times of the closest optical peaks to the expected outburst are indicated at the top of each figure.}
    \label{fig:LC}
\end{figure*}

\begin{figure}
\includegraphics[width=\linewidth, trim=0 0 0 0, clip]{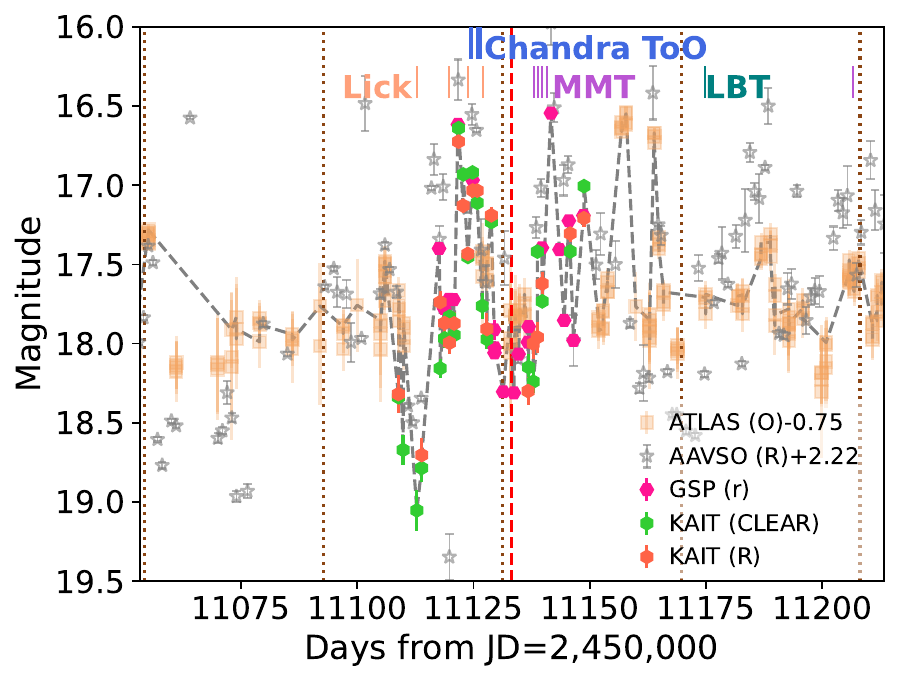}
\caption{AT~2016blu was monitored with KAIT, GSP, and AAVSO observers in March 2026 as it approached its expected outburst. When it reached a peak magnitude of 16.6, \chandra\ ToO observations were triggered, resulting in a total exposure of $\sim$ 100 ks obtained over five epochs. Optical spectra were obtained during this period with Lick/Kast and later with MMT/BCH and LBT/MODS  (see Figure~\ref{fig:spectra}).
\label{fig:LCchandra}}
\end{figure}

Figure~\ref{fig:LC} shows that AT~2016blu underwent five additional outbursts between January 2024 and the sixth outburst, which coincided with the \chandra ToO observations; see Figure~\ref{fig:LCchandra}. AT~2016blu exhibits multiple optical peaks of comparable magnitude within a single event. The timing of some peaks is consistent with a period of $\sim$ 113 days (shown with red dashed lines), while some others align with a shorter period of $\sim$ 38.5 days (shown with brown dashed lines); see \cite{A2316blu, A25} for further details. Individual peaks defined by one data point, such as the one in 2024 February, should be interpreted with caution, since as described above the datasets were obtained with different telescopes, filters, and reduction procedures; some systematic scatter between them is expected.

To investigate the color evolution of AT~2016blu during the 2026 outburst, we compute color indices using the multiband KAIT photometry. Since the observations in different filters are not strictly simultaneous, we linearly interpolate the $B$, $V$, and $I$  light curves onto the $R$-band epochs, keeping only interpolated points within 5 days of an observation. Associated uncertainties were propagated in quadrature from the individual band uncertainties. See Section~\ref{sec:visual_results} for a detailed discussion of the color evolution of AT~2016blu.

\subsection{Optical Spectroscopy}\label{sec:obsspec}

\begin{figure*}[t]
  \centering
  \includegraphics[width=0.9\textwidth, trim=0 0 0 0, clip]{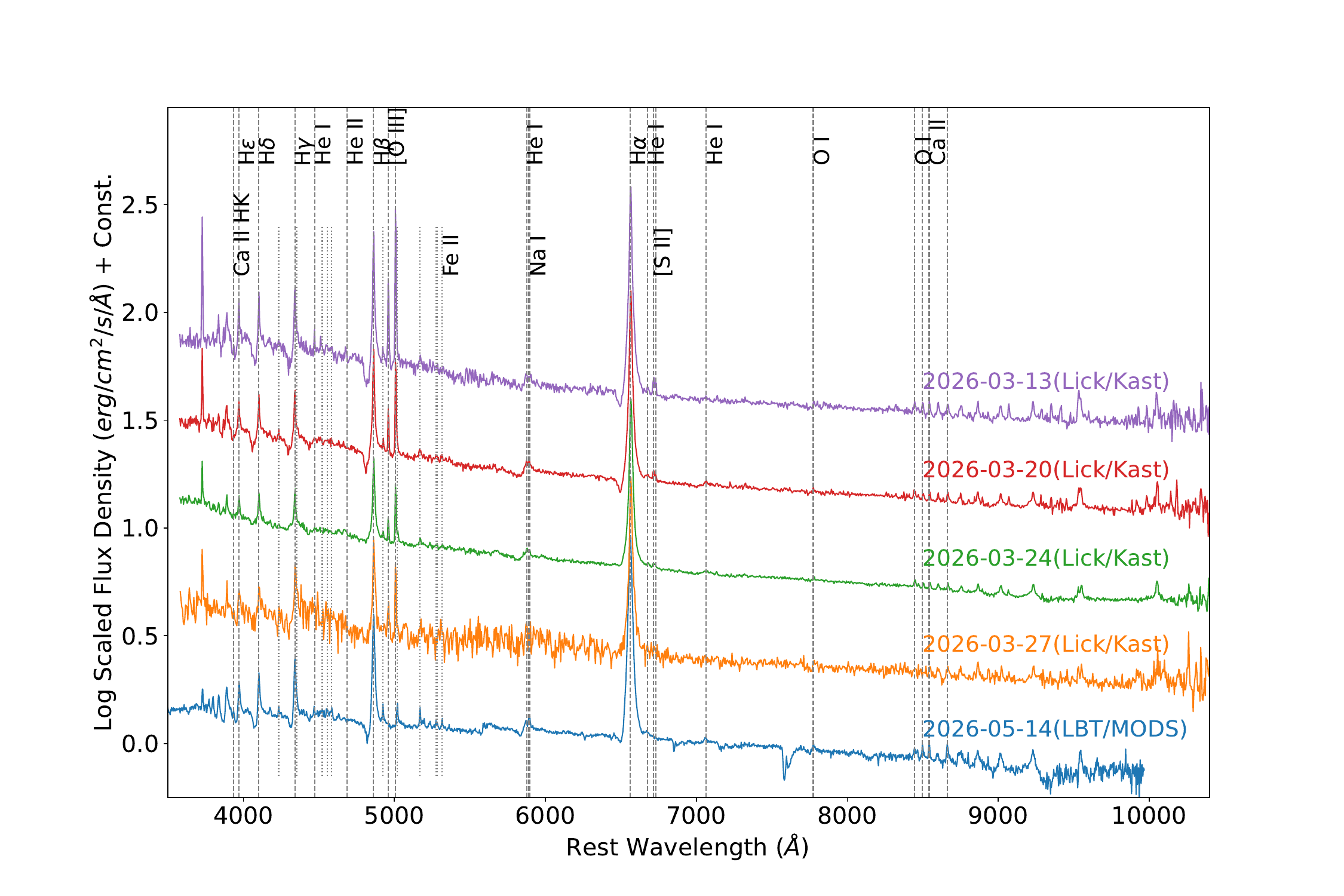}\\
  \vspace{0.5em}
  \includegraphics[width=0.9\textwidth, trim=0 0 0 80, clip]{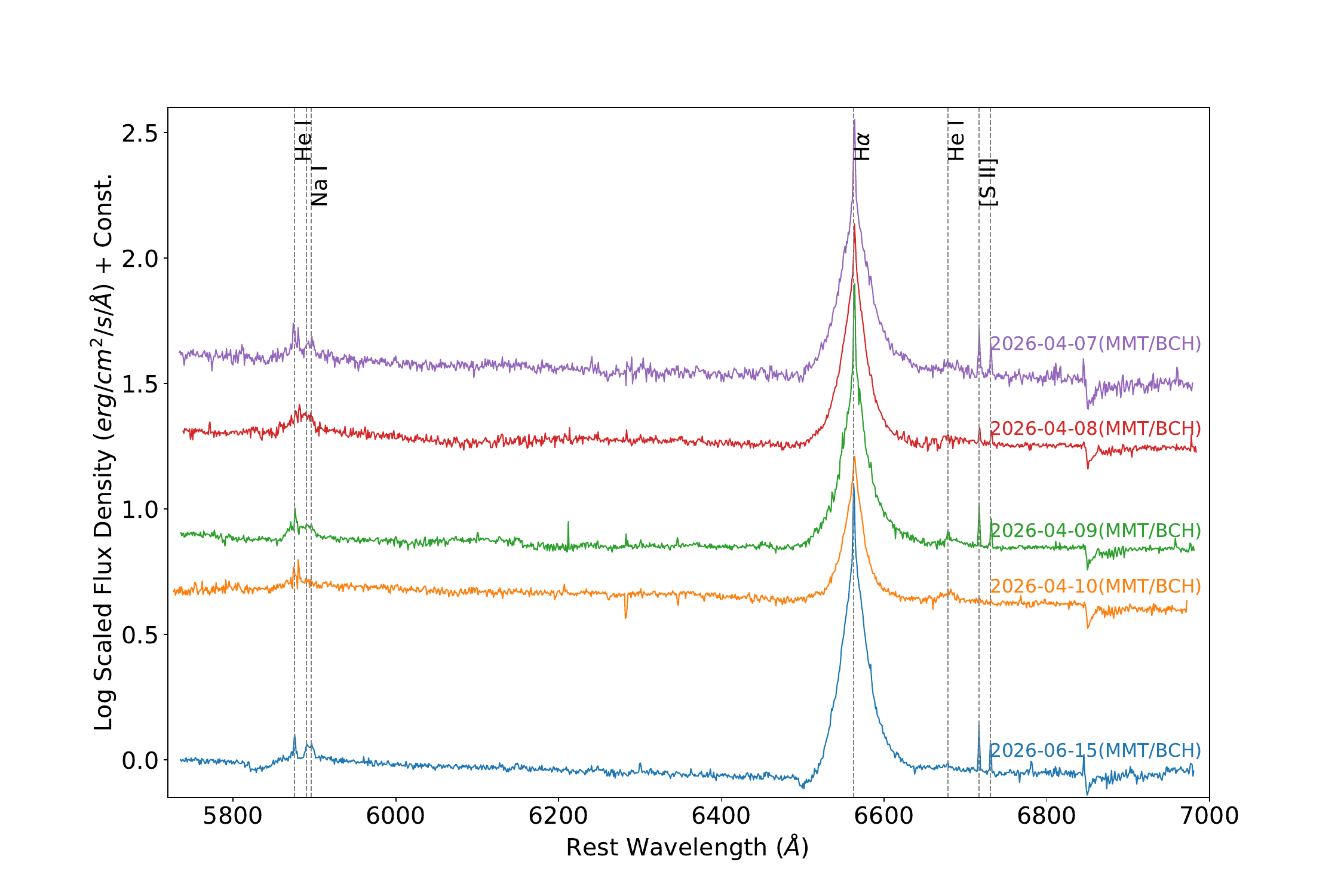}
  \caption{The top panel shows the optical spectra of AT~2016blu obtained with Lick/Kast during its March 2026 outburst, including some epochs obtained during the \chandra\ ToO observations (see Figure~\ref{fig:LCchandra}) and a later epoch obtained with LBT/MODS. The spectra exhibit H$\alpha$, He~{\sc i}, O~{\sc i}, Ca~{\sc ii}, and [O~{\sc iii}] lines. A P~Cygni profile in H$\alpha$ is present during the quiescent phase and at the optical peak, and becomes less prominent shortly thereafter as the system fades, while still remaining in the outburst phase. The bottom panel shows optical spectra obtained with MMT/BlueChannel using the 1200 lines mm$^{-1}$ grating prior to the second optical peak in April and later in June during a fainter optical peak. These spectra display He~{\sc i} emission lines but no obvious P~Cygni profile in H$\alpha$ until the final epoch obtained in June 2026. The dip near 6860\,\AA\ is telluric absorption (B band).}
  \label{fig:spectra}
\end{figure*}

Optical spectra of AT~2016blu were obtained during its March 2026 outburst, some during the \chandra\ ToO observations (see Figure~\ref{fig:LCchandra}), using the Kast double spectrograph on the Lick 3~m Shane reflector \citep{Miller1993}. AT~2016blu spectra were also obtained using the BlueChannel (BCH) Spectrograph on the 6.5~m Multiple Mirror Telescope (MMT) with a 1200 lines mm$^{-1}$ grating, prior to the second optical peak in April. Additional spectra were obtained in May and June using the Multi-Object
Double Spectrograph (MODS) on
the 8 m Large Binocular Telescope \cite[LBT;][]{B00} and MMT/BCH. Figure~\ref{fig:spectra} presents the spectra, corrected for a redshift of $z = 0.00261$. The UTC dates of the observations are indicated alongside each spectrum; see Figure~\ref{fig:spectra}.

\begin{figure}
\includegraphics[width=\linewidth, trim=0 0 0 0, clip]{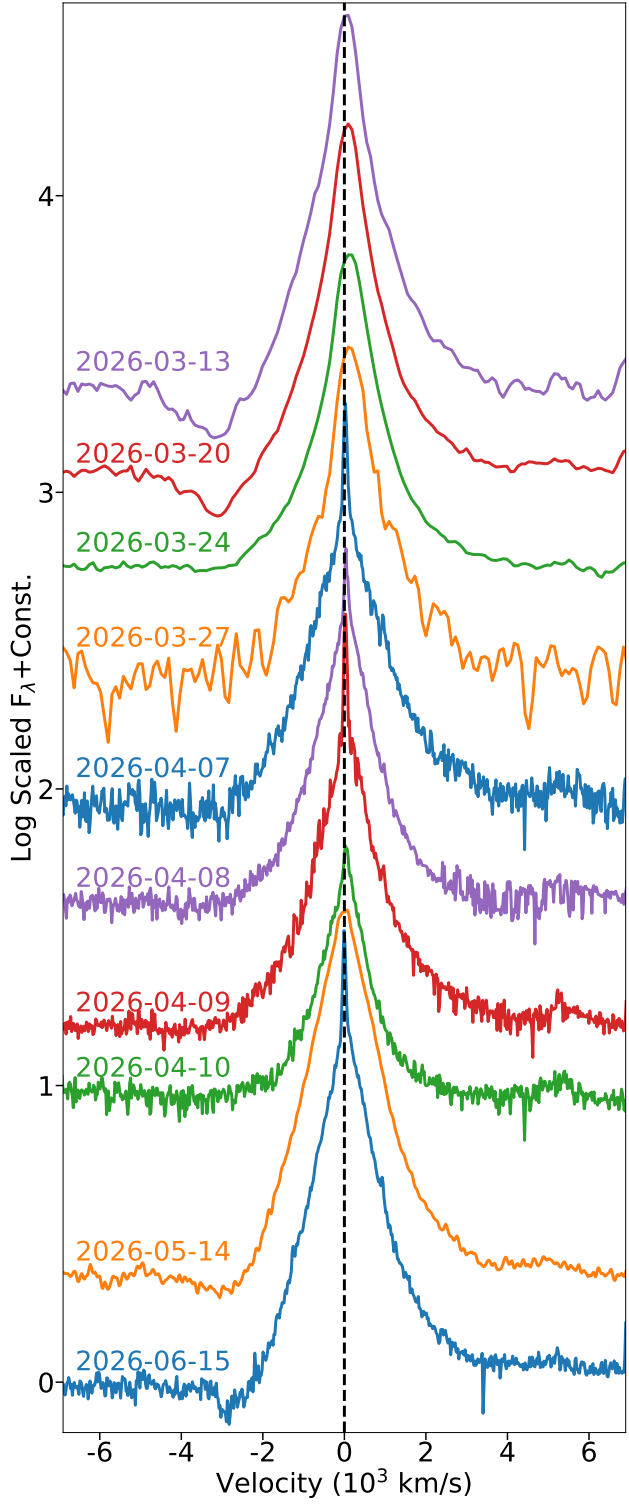}
\caption{Evolution of the H$\alpha$ profile across ten epochs from 2026 March to June, shown in velocity space relative to the rest wavelength of H$\alpha$. The P~Cygni profile is present during quiescence and at the optical peak, weakens as the system fades, disappears before the second optical peak in April, and reappears in May.}
\label{fig:AllHalpha}
\end{figure}

AT~2016blu exhibits H$\alpha$ and He~{\sc i} lines, indicating a hot LBV, similar to the progenitor of SN~2009ip \citep{S10}. Its spectrum also shows O~{\sc i}, Ca~{\sc ii}, and [O~{\sc iii}] features. He~{\sc ii} $\lambda$4686 is not detected in the spectra. In the first spectrum, there is a tentative feature near this wavelength, though it is weak and not unambiguously identified above the continuum noise level. Figure~\ref{fig:AllHalpha} shows that the P~Cygni profile in the H$\alpha$ line is present during the quiescent phase and at the optical peak, and becomes less prominent shortly thereafter as the system fades, while still remaining in the outburst phase. Prior to the second optical peak in April, the P~Cygni profile in H$\alpha$ is no longer detected, and reappears later in the last two spectra obtained in May and June.
LBVs in outburst typically exhibit P~Cygni profiles and are relatively cool, which is not the case for AT~2016blu.
See \cite{A25} for the detailed data reduction and analysis of the AT~2016blu spectra.

\section{Results}\label{sec:result}
\subsection{Visual-Wavelength Variability During the X-ray Observations}\label{sec:visual_results}

Figure~\ref{fig:LCchandra} presents the optical light curve of AT~2016blu in March 2026 alongside the \chandra\ observations.  AT~2016blu  brightened by more than 1 mag in March 22. When it reached 16.6 mag, \chandra\ ToO observations were triggered, primarily based on KAIT and GSP data. The X-ray results are discussed in the next section. In April and May, AT~2016blu exhibited additional optical peaks, consistent with the multipeaked behavior of comparable magnitude reported in previous outbursts.

\begin{figure}
\includegraphics[width=\linewidth, trim=0 0 0 0, clip]{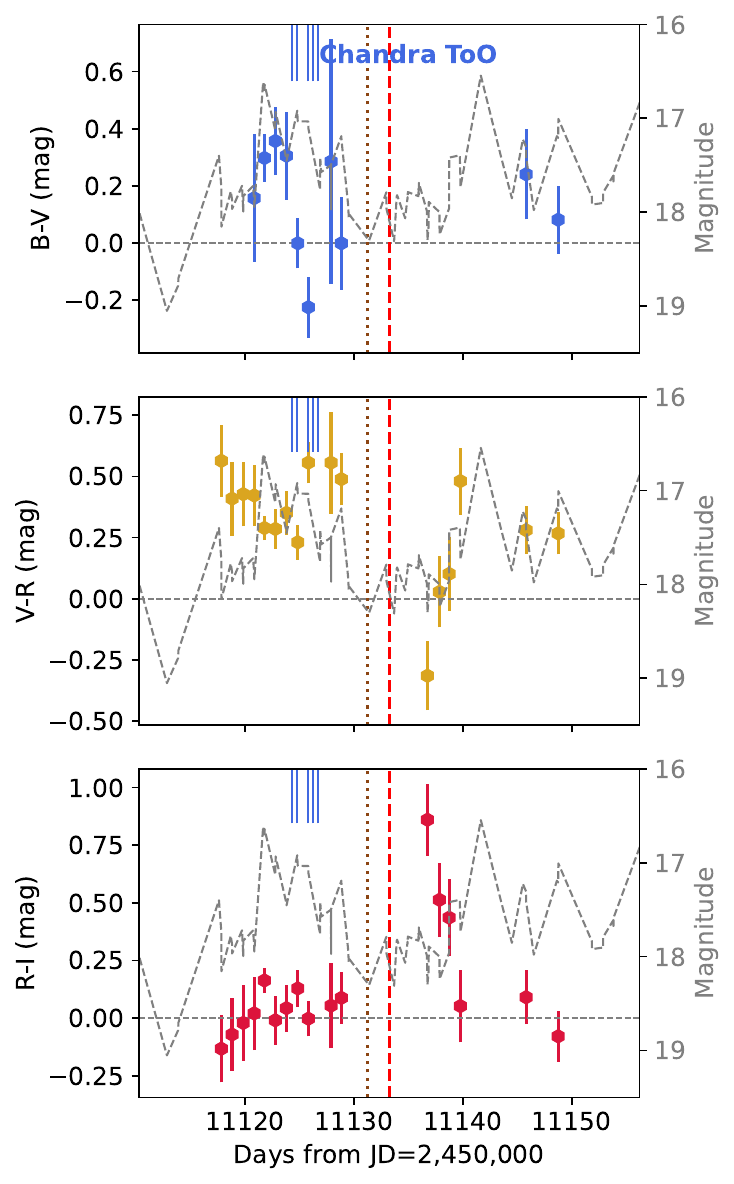}
\caption{Color evolution of AT~2016blu during its 2026 outburst. The data are obtained from Lick/KAIT, with magnitudes interpolated. The gray dashed line corresponds to that shown in Figure \ref{fig:LCchandra}, representing the combined magnitude evolution from multiple telescopes. An anticorrelated evolution is observed between $V-R$ and $R-I$: $V-R$ transitions from red to blue and then reddens again, while $R-I$ reddens from an initial value near zero before returning to nearly zero. 
\label{fig:color}}
\end{figure}

Figure~\ref{fig:color} shows the color evolution of AT~2016blu using KAIT data.  The light curve constructed by combining data from multiple telescopes is overlaid on each color panel as a gray dashed line to facilitate comparison between the photometric state of the source and its color evolution. A clear anticorrelation is seen between $V-R$ and $R-I$, such that $V-R$ evolves from red to blue and then reddens again, while $R-I$ shows the opposite trend, reddening from an initial value near zero before returning to approximately zero. The anticorrelation might be due to variation in H$\alpha$ emission. During the \chandra\ ToO observations, $B-V$ becomes bluer while $V-R$ becomes redder, and $R-I$ becomes bluer; however, we note that the ToO color evolution is based on only two data points. 

\subsection{X-ray Luminosity and Variability}\label{sec:x-ray_results}

\begin{figure}
\includegraphics[width=\linewidth, trim=0 0 0 0, clip]{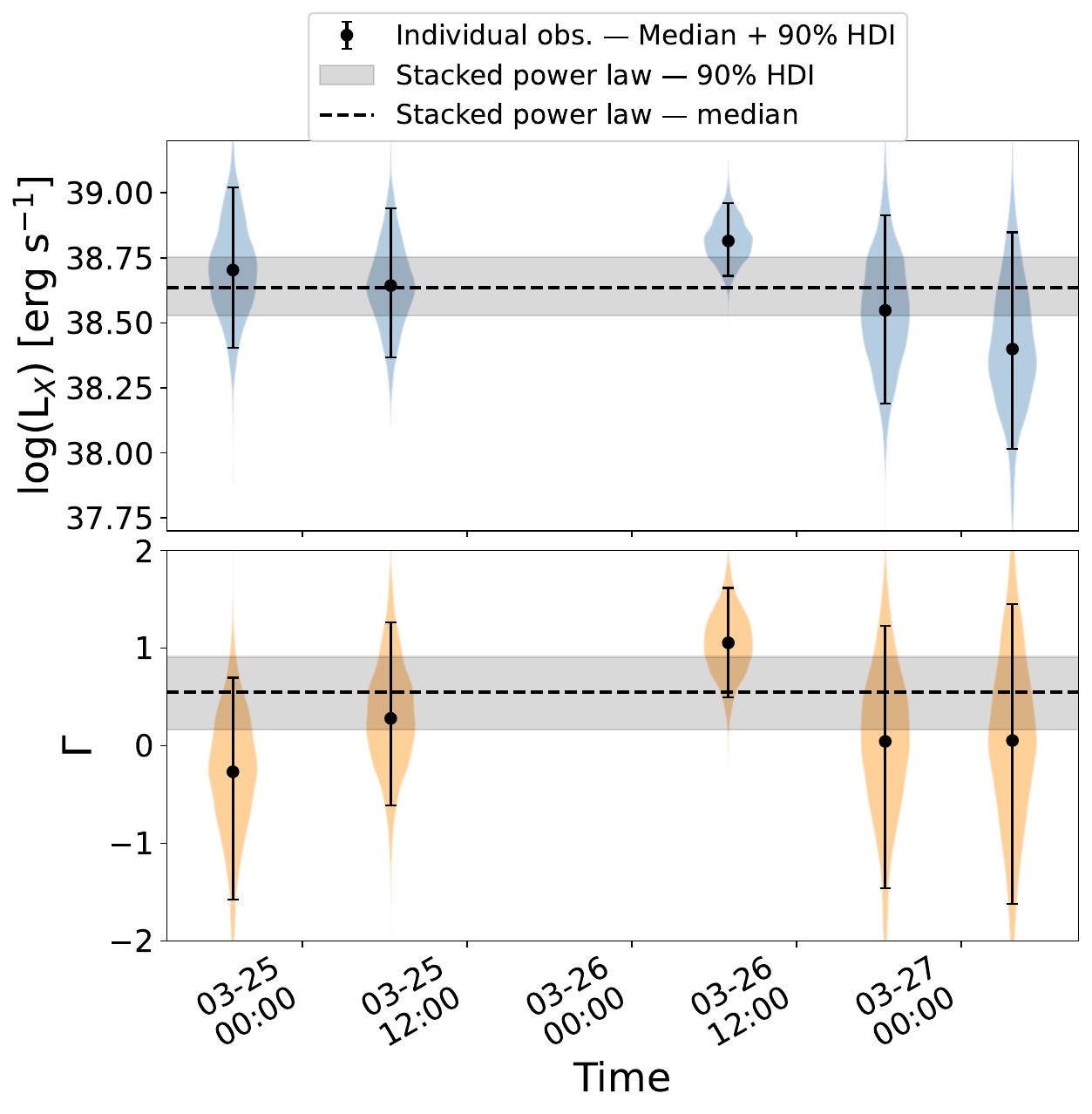}
\caption{X-ray luminosity (top) and photon index (bottom) derived from the X-ray fit of each ToO observation listed in Table~\ref{tab:chandra_obs}. The full posterior is shown as a distribution, while the points represent the median, and the error bars correspond to the 90\% HDI. The X-ray luminosity remains consistent across all five epochs within the uncertainties. The shaded area corresponds to the results of the fit to a stacking of all observations, using a simple power-law model. 
\label{fig:xray_fit_results}}
\end{figure}

Using BXA as described in Section~\ref{sec:x-ray_obs}, we obtain an estimate for the X-ray luminosity and power-law index ($\Gamma$) of AT~2016blu for each individual observation. The results are summarized in Table~\ref{tab:xray_fit_results} and shown in Figure~\ref{fig:xray_fit_results}. For all parameters except $N_{\rm H}$, we report the median value of the posterior, as well as the 90\% highest-density interval (HDI; i.e., 90\% range of the posterior). For $N_{\rm H}$, we report upper limits as the 95\% range of the posterior. 

\begin{table}[h!]
    \centering
    \resizebox{\columnwidth}{!}{%
    \begin{threeparttable}
    \caption{\chandra X-ray Fit Properties of AT~2016blu.}
    \label{tab:xray_fit_results}
    \begin{tabular}{ccccc}
        \toprule
        \textbf{ObsID} & \textbf{Date} &\textbf{log L$_{\rm x}$} & \textbf{$\Gamma$} & \textbf{log $N_{\rm H}$}\\
         & (UTC) & (erg~s$^{-1}$) & (cm$^{-2}$)\\
        \midrule
        31041 & 2026-03-24 & 38.70$^{+0.32}_{-0.29}$ & -0.27$^{+0.96}_{-1.31}$ & $-$\\
        32254 & 2026-03-25 & 38.64$^{+0.30}_{-0.28}$ & 0.28$^{+0.99}_{-0.89}$ & $-$\\
        32255 & 2026-03-26 & 38.81$^{+0.15}_{-0.13}$ & 1.05$^{+0.56}_{-0.56}$& $-$\\
        32256 & 2026-03-26 & 38.55$^{+0.37}_{-0.36}$ & 0.04$^{+1.18}_{-1.50}$& $-$\\
        32257 & 2026-03-27 & 38.40$^{+0.45}_{-0.38}$ & 0.05$^{+1.40}_{-1.67}$& $-$ \\
        \midrule
        \multicolumn{5}{c}{Power law} \\
        \midrule
        stacked & $-$ & 38.64$^{+0.12}_{-0.11}$ & 0.55$^{+0.36}_{-0.38}$ & $-$\\
        \midrule
        \multicolumn{5}{c}{Abs. Power law} \\
        \midrule
        stacked & $-$ & 38.64$^{+0.11}_{-0.11}$ & 0.66$^{+0.52}_{-0.41}$ & 21.87$^{u}$ \\
        \bottomrule
    \end{tabular}
    \begin{tablenotes}
        \footnotesize
        \item[u] Upper limit.
    \end{tablenotes}
    \end{threeparttable}}
\end{table}




Despite the uneven number of counts per observation, as shown in Table~\ref{tab:chandra_obs}, the luminosities of each individual observation remain consistent within the uncertainties, in the range of ${\log L_{\rm X}}\approx38.7$~erg~s$^{-1}$. Considering the full error range, the source is barely consistent with breaching the ULX threshold at ${\log L_{\rm X}}\approx39$~erg~s$^{-1}$. 

\begin{figure*}
\includegraphics[width=0.495\linewidth, trim=0 0 0 0, clip]{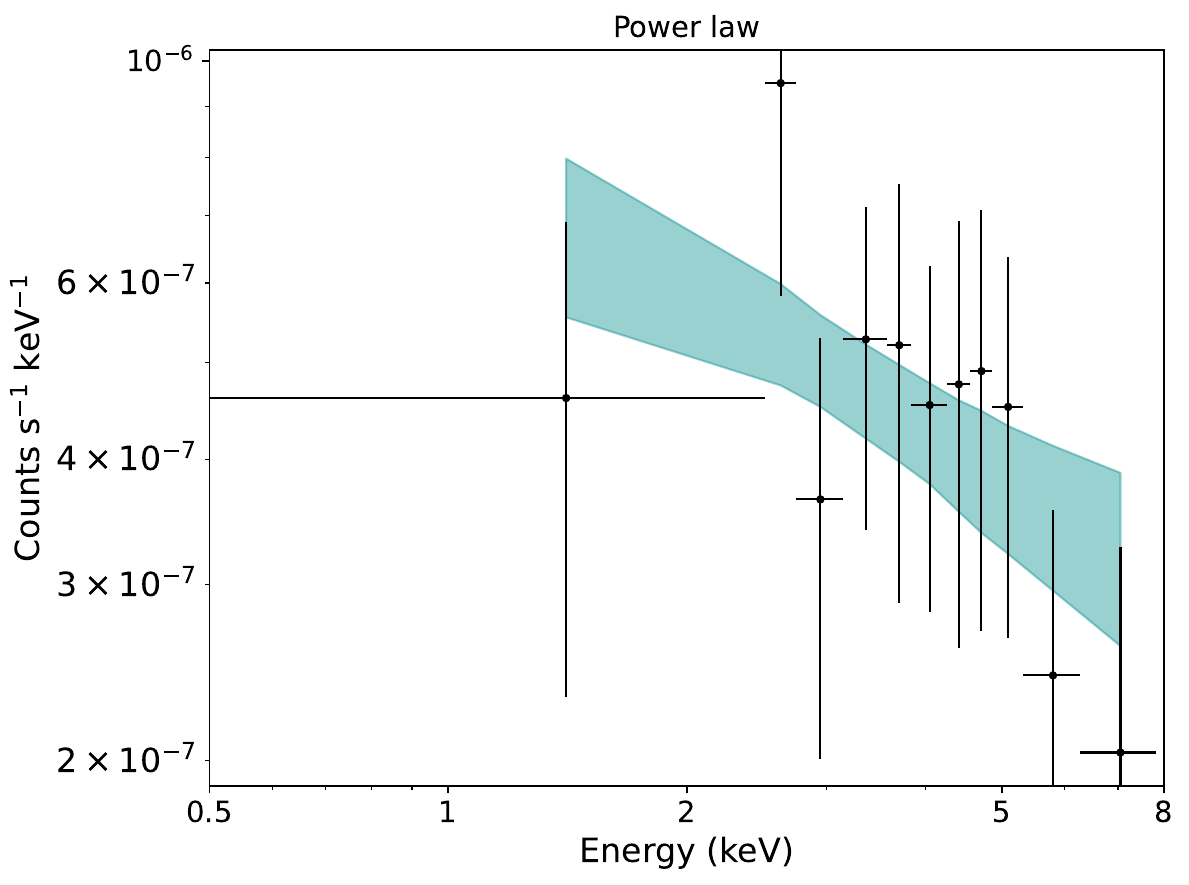}
\includegraphics[width=0.495\linewidth, trim=0 0 0 0, clip]{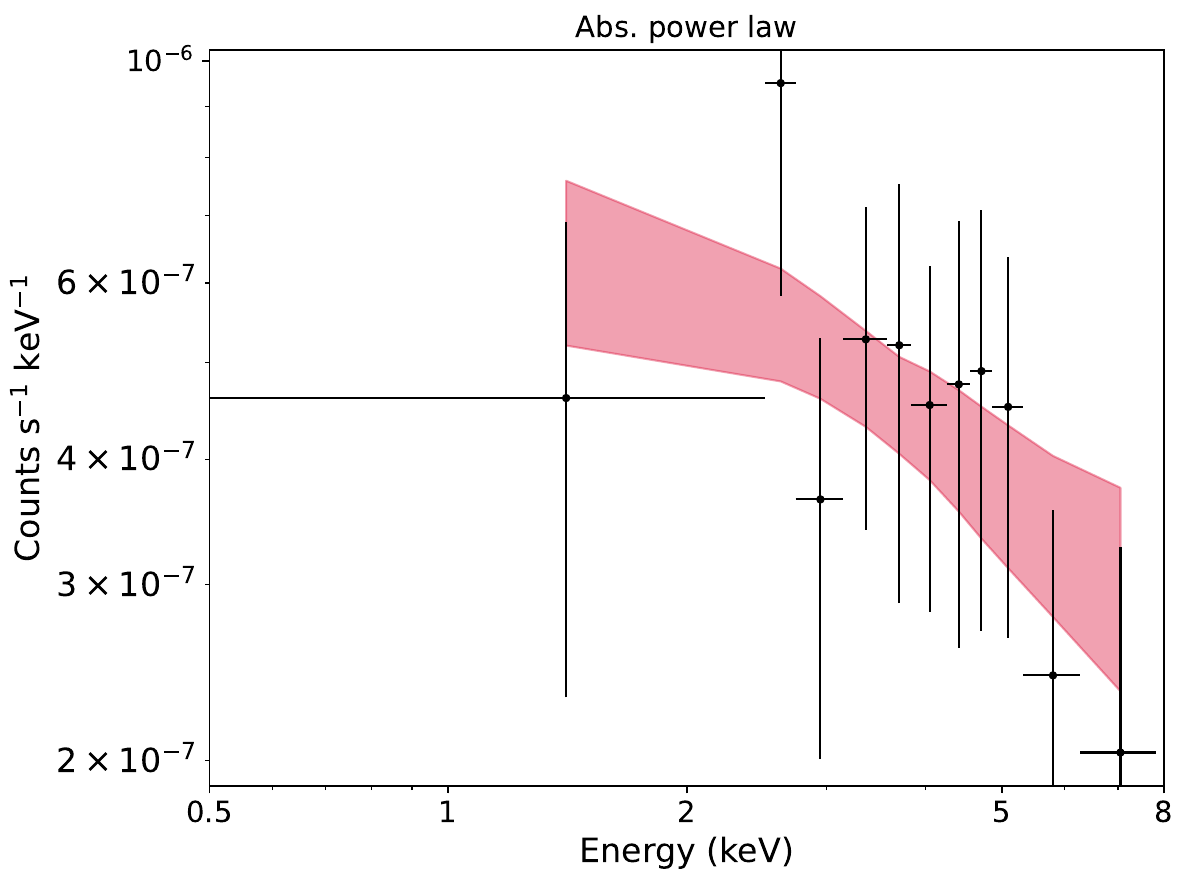}
\caption{Fits to the stacked X-ray spectra using a power-law (left) and absorbed power-law (right) model. The shaded bands correspond to the 68\% range of the posterior.  Although stacking improves the constraints on the luminosity, the limited number of counts still results in substantial uncertainty in the spectral shape parameters.
\label{fig:xray_spectra}}
\end{figure*}

Given the lack of obvious variability in the closely separated observations, we also provide the results of fitting the stacked data. For both models considered, the luminosity determination is almost identical, at ${\log L_{\rm X}}\approx38.64 \pm 0.11$~erg~s$^{-1}$. Owing to the similarity of both model fits, we opt to show only the results of the simple power-law fit in Figure~\ref{fig:xray_fit_results}. With the tighter constraints, the source does not quite reach the ULX threshold within errors, but falls only $\sim0.2$~dex below. Even when allowing for obscuration, the preferred value of $N_{\rm H}$ is low, resulting in a very hard spectrum, with a low photon index value, in the range of $\Gamma\approx0.2$--1.2. The stacked spectrum, along with the 68\% range of the posterior, are shown in Figure~\ref{fig:xray_spectra}.

\section{Discussion} \label{sec:discussion}

\subsection{Association Between the X-ray Detection and AT 2016blu}

As shown in Figure~\ref{fig:chandra_and_xmm}, the stacked \textit{Chandra} emission falls within the $2''$ region centered on the TNS optical position of AT~2016blu (magenta circle); had we instead adopted the SIMBAD position, which is offset by $\sim 1''$, the X-ray centroid would lie off-center. 
\chandra's absolute pointing accuracy is generally better than $0.4''$\footnote{https://cxc.cfa.harvard.edu/ciao/threads/reproject\_aspect}, and the absolute astrometric accuracy of the source positions in the \chandra Source Catalog \cite[CSC;][]{Evans2024} has been found to be typically $\sim0.16''$ \citep{Rots2011}. 

\begin{figure*}
    \centering
    \includegraphics[width=0.95\textwidth, trim=0 0 0 0, clip]{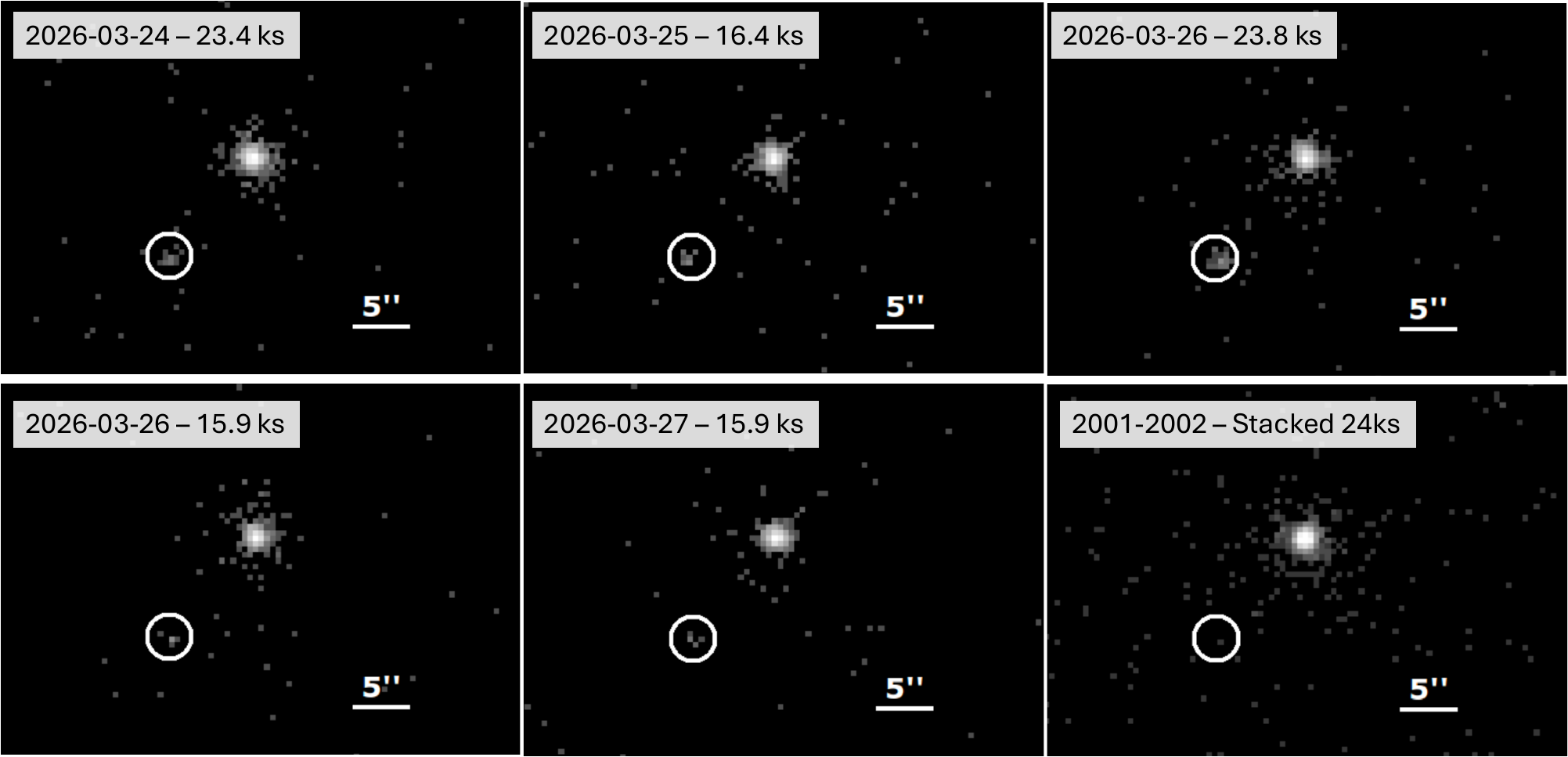}
    \caption{Comparison between all \chandra observations of AT~2016blu during the ToO campaign, and a stacked image of previous archival observations. The white circle, of $2''$ radius, marks the optical position of AT~2016blu. No source is detected in the stacked archival images prior to the first observed optical outburst of AT~2016blu, when it was a stable system.}
    \label{fig:chandra_all_obs}
\end{figure*}

The alternative to a genuine counterpart to AT~2016blu would be a background quasar with a coincident position. This quasar, furthermore, would have brightened at a time coincident with our observations, and thus the AT~2016blu outburst. Indeed, a stacking of the available \chandra data before its first outburst in 2012 (see Table~\ref{tab:chandra_obs}) results in a total observing time of $\sim$24~ks, comparable with our longest observations. No source compatible with the position of AT~2016blu is detected in the stacked image. However, we detect the source in each of our observations individually, even the shorter ones ($\sim16$~ks). Furthermore, \chandra's ACIS-S detectors were more sensitive in 2001$-$2002, making the stacked archival observation even deeper in comparison with the longest exposures of our 2026 ToO, as illustrated in Figure \ref{fig:chandra_all_obs}.

We also estimate the likelihood of a background quasar at the observed source brightness (or higher) being inside the $2''$ radius around AT~2016blu. The flux of the detected source is $\sim3 \times 10^{-14}$~erg~cm$^{-2}$~s$^{-1}$. We compare this flux against the 2--10~keV log(N) $-$ log(S) distribution of \cite{Ueda2014}, which yields a density of active galactic nuclei (AGNs) at or above the given flux threshold of $\sim57.7$/deg$^2$. A radius of $2''$ corresponds to an area of 9.69$\times10^{-7}$~deg$^2$. Therefore, the expected number of background AGNs with this given brightness within the uncertainty region is $\sim6\times10^{-5}$. It is therefore extremely unlikely that the detected X-ray source is a background AGN rather than the genuine X-ray counterpart of AT~2016blu.

We assume, therefore, that the X-ray source is associated to AT~2016blu. This allows us to obtain a luminosity upper limit for the nondetection in the archival data, assuming (1) no variability between the three different observations, and (2) taking the sensitivity of the 2002 observation as representative of the whole period\footnote{This is a reasonable assumption, given how the observations were taken roughly within a year of each other. Regardless, we use the latest (less sensitive) observation, in order to produce a more conservative upper limit.}. Using the response files associated to that observation, we simulate 1000 fake spectra using the \texttt{Xspec} function \texttt{fakeit} and measure the number of counts produced. We start at the luminosity measured for the stacked observation (log $L_{\rm X}\approx38.6$~erg s$^{-1}$), which generates a median of 36~counts. We decrease the luminosity log $L_{\rm X} = 0.1$ until 95\% of the simulated observations produce 1 count or less (a single count is detected in the stacked observation). This results in an upper limit for the nondetection of log $L_{\rm X} < 36.6$~erg s$^{-1}$. Comparing this upper limit with the 2026 X-ray detection indicates that the X-ray luminosity of AT~2016blu increased by a factor of $\sim 100$.

\subsection{Physical Origin of the X-ray Emission}
Four scenarios are commonly invoked to potentially account for X-ray emission from LBVs. (1) Intrinsic shocks in the wind, which typically produce relatively soft X-ray spectra with luminosities below $\sim10^{31}$~erg~s$^{-1}$ \citep{Naze12}. Clearly this scenario cannot explain the $>$10$^7$ times more luminous X-rays in AT~2016blu. Moreover, the light curve and spectra of AT~2016blu already indicate that it is not a single star, and the X-ray observations further support this interpretation. (2) Colliding winds, if the companion is another massive star, which yield moderately high X-ray luminosities in the range $\sim10^{31}$--$10^{34}$~erg~s$^{-1}$ \citep{Naze12}. Although the quasiperiodicity of AT~2016blu  points to a binary system, massive star colliding-wind X-ray luminosities are still 10,000 times too faint to explain AT~2016blu X-ray luminosity. (3) Collisions between successive ejected shells. Material ejected at
consecutive periastron passages could collide and shock heat the gas. However, for a strong shock the post-shock temperature given by the Rankine-Hugoniot jump conditions depends on the
shock velocity,
$kT \approx (3/16)\,\mu\, m_{\rm H}\, v_s^{2} \approx 0.3\,
( v_s / 500\,\mathrm{km\,s^{-1}})^{2}\ \mathrm{keV}$,  where $\mu = 0.6$ is the mean molecular
weight of fully ionized gas of solar composition. For colliding shells, the shock velocity $v_s \approx \Delta v$, the velocity difference between them, with the exact factor depending on their density contrast. Since $\Delta v$ cannot exceed the
spread in ejection velocities, which the H$\alpha$ profiles limit to a few hundred $\mathrm{km\,s^{-1}}$, such a collision yields a soft X-ray spectrum, inconsistent with the hard spectrum we measure. (4) Accretion onto a compact companion, which can produce strong, hard X-ray emission with luminosities substantially higher ($\gtrsim10^{37}$~erg~s$^{-1}$; \citealt{Binder2016}). As described above, AT~2016blu exhibits $\log L_{\rm X} \approx 38.64$~erg~s$^{-1}$, only achievable by accretion onto a compact companion. We therefore conclude that its outbursts are most likely driven by accretion onto such an object. Since the system consists of a blue straggler (an LBV with a mass of at least 33 $M_\odot$; \citealt{A2316blu}) and a compact object, AT~2016blu can be classified as a high-mass X-ray binary (HMXB). 

Although the high X-ray luminosity implies accretion onto a compact object, distinguishing whether the accretor is a neutron star or a black hole is not possible for most sources. That is mainly because both source classes can produce hard, power-law-like X-ray spectra with comparable luminosities, and their spectral and timing properties are generally indistinguishable \citep[e.g.,][]{DiSalvo2006,Munoz-Darias2014}. A dynamical mass would distinguish the two, but the H$\alpha$ radial velocities do not track the orbital phase and therefore do not yield an orbital solution for the companion mass \citep{A25}. In a few instances, coherent pulsations \citep[e.g.,][]{Bachetti2014,Furst2016,Israel2017,Kosec2018,Rodriguez-Castillo2020}, as in SN~2010da (see Section~\ref{sec:comparison}), or type I X-ray bursts\footnote{A thermonuclear explosion on the surface of an accreting neutron star. As this process requires a physical surface, it cannot originate in a black hole.} \citep[e.g.,][]{Galloway2021} can be uniquely associated with neutron star accretors.  In the particular instance of AT~2016blu, the \chandra data are insufficient to search for pulsations (which would require a binning of seconds -- not possible with limited counts);  \xmm data, despite the larger sensitivity, would be dominated by the nearby ULX, NGC~4559~X7. A type I X-ray burst typically lasts for a few tens of seconds, or up to hours, in the case of a superburst \citep{Parikh2013}; and therefore does not match the data presented here. In conclusion, as with the majority of HMXBs, we cannot firmly associate AT~2016blu with a compact object type.

It is worth mentioning that, while AT~2016blu does not meet the threshold for a ULX classification, its very high brightness likely places it in the same source class. In particular, its X-ray luminosity corresponds to the Eddington limit for a $\sim$3.5~M$_\odot$ compact object, making it super-Eddington for a neutron star. Furthermore, given the short time separation between our observations, we cannot ascertain when the luminosity maximum may have occurred, and how much higher it would be; potentially crossing into the ULX threshold. As such, we call AT~2016blu a ULX candidate. 

Finally, we consider  possible reasons why AT~2016blu was not detected in the 2001--2002 archival data but became active after 2012. The pre-2012 KAIT monitoring shows
that the system was stable, with no optical outbursts detected
\citep{A2316blu}; the archival nondetection thus appears to reflect
a quiescent state of the system. One possibility is that the primary
expanded as a result of its evolution, as suggested by the rise in the
\textit{Gaia} baseline \citep{A2316blu}, and that its
larger radius allowed it to begin interacting with the companion, initiating
the eruptions in 2012. Alternatively, a known property of LBVs is that their radii increase during S Dor phases on timescales of decades. If so, the companion could have remained in a quiescent accretion state while the LBV had a smaller radius, with the interaction and accretion rate increasing once the primary expanded during an S Dor phase. A prediction of this scenario is that the activity should eventually stop if the S Dor phase ends and the LBV returns to its quiescent, smaller-radius state. So far, however, the system has remained active since turning on, which favors the long-term evolutionary expansion. 

\subsection{Comparison with Other LBVs and SN Impostors}\label{sec:comparison}

\begin{figure}
\includegraphics[width=\linewidth, trim=0 0 0 0, clip]{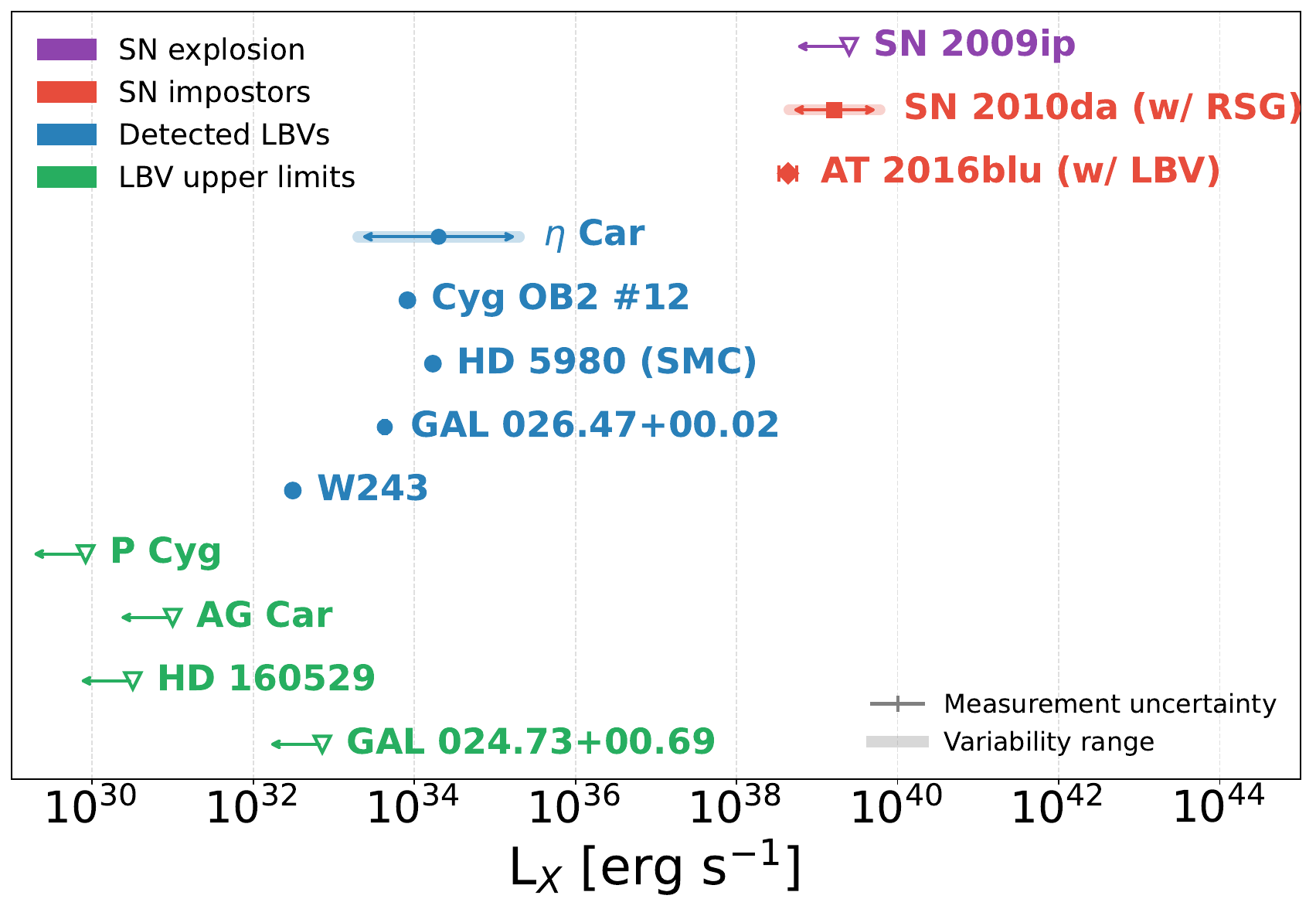}
\caption{AT~2016blu’s X-ray luminosity is shown in comparison with examples of other LBVs, SN impostors, and the SN explosion of SN~2009ip. AT~2016blu exhibits X-ray luminosities significantly higher than those of LBVs with colliding winds, and it is the first LBV showing  X-ray luminosity consistent with accretion onto a compact object.}
\label{fig:xray_luminosities_comparison}
\end{figure}

Figure \ref{fig:xray_luminosities_comparison} compares the X-ray luminosity of AT~2016blu with that of known LBVs and SN impostors, all discussed in detail below. AT~2016blu exhibits significantly higher X-ray luminosities than those of any previously studied LBV and is more comparable to SN impostors, making it the first LBV with X-ray luminosity consistent with accretion onto a compact object. Considering the error bars, the X-ray luminosity of AT~2016blu is also comparable to that observed in SN~2009ip ($L_{\rm X} \leq 2.5 \times 10^{39}$\,erg\,s$^{-1}$), measured near the peak of its 2012b explosion, although in that case the emission arises from the SN explosion interacting with CSM rather than the LBV progenitor's eruption \citep{Margutti14}. Below we summarize the X-ray results from LBVs and an SN impostor, SN~2010da.

LBVs: $\eta$~Car was detected in X-rays 
\citep{Seward1979, Corcoran2010}, with luminosity 
varying from $\sim$$2 \times 10^{35}$\,erg\,s$^{-1}$ at maximum to 50--100 
times smaller at minimum, attributed to colliding winds in an eccentric binary system \citep{C97, P02, C05, P09, P11, R16}. Note that the minimum-to-maximum X-ray luminosity shown in Figure \ref{fig:xray_luminosities_comparison} is detected over 100 yr after the Great Eruption. The very fast and strong wind inferred from models of the X-ray emission indicates that $\eta$ Car's companion star is most likely a W-R star \citep{smith18}. 
Similarly, Cyg~OB2~\#12 was detected with $L_{\rm X} \approx 8.2 \times 10^{33}$\,erg\,s$^{-1}$ \citep{Harnden1979, Rauw2011, Naze12}. Subsequent high-resolution \chandra spectroscopy of Cyg~OB2~\#12 
confirmed its colliding wind binary nature \citep{Oskinova2017}.
The SMC LBV HD~5980, which is a quadruple system that harbors a W-R star, was also studied in X-rays, 
reaching a luminosity of $\sim 1.7 \times 10^{34}$\,erg\,s$^{-1}$, consistent with colliding winds \citep{Naze2002}. In contrast, P~Cyg was 
observed with {\it ROSAT} but remained undetected, yielding only an upper limit of 
$\sim 10^{31}$\,erg\,s$^{-1}$ \citep{Berghofer2000}.

The X-ray properties of Galactic LBVs were then systematically studied by 
\citet{Naze12} using \textit{XMM-Newton} and \textit{Chandra}. Several 
well-studied LBVs such as P~Cyg, AG~Car, and HD~160529 (see \citealt{Naze12} for additional examples) were undetected, with 
upper limits reaching as low as $L_{\rm X} \approx 8.3 \times 10^{29}$\,erg\,s$^{-1}$. 
Only two objects were detected: GAL~026.47+00.02 
($L_{\rm X} \approx (4.3 \pm 0.2) \times 10^{33}$\,erg\,s$^{-1}$), and W243 
($L_{\rm X} \approx 1.5$--$3.1 \times 10^{32}$\,erg\,s$^{-1}$), the bright hard 
X-ray emission of the first one was attributed to wind-wind collisions in binary 
systems.  
The intrinsic 
X-ray faintness of LBVs in \citet{Naze12} was attributed to their slow winds, which are 
insufficient to drive strong internal shocks, and/or to heavy wind absorption. 
Importantly, \citet{Naze12} noted that the absence of X-ray emission cannot be 
used to infer a single-star nature, as even X-ray bright colliding-wind binaries 
are the exception rather than the rule among massive binaries. Therefore, in summary, no previous X-ray studies of LBVs have reported luminosities as high as that observed for AT~2016blu.

SN impostor SN~2010da: an optical transient in NGC~300 \citep{M10}, it was associated with a bright \textit{Swift}/XRT X-ray source ($\sim4.5^{+0.9}_{-2.1}\times10^{38}$\,erg\,s$^{-1}$; \citealt{Immler2010, Binder2011}). A \chandra\ observation later showed the source had faded but remained too luminous for a single massive star outburst, indicating a HMXB nature \citep{Binder2011}. Subsequent \chandra\ observations revealed recurring X-ray outbursts, consistent with accretion onto a neutron star in a HMXB \citep{Binder2016}. Multiwavelength follow-up observations by \citet{Villar2016} detected an ultraluminous outburst ($L_{\rm X} \approx 6\times10^{39}$\,erg\,s$^{-1}$) during the 2010 event, and detected persistent He~{\sc ii} $\lambda4686$ emission and late-time coronal iron lines attributed to a compact companion, though the nature of the compact object remained uncertain. Furthermore, \citet{Villar2016} argued that the primary is not an LBV but rather a yellow supergiant. Mid-infrared rebrightening and dust properties consistent with a sgB[e] star led \citet{Lau2016} to propose a sgB[e]–HMXB classification, while VLT/X-Shooter spectroscopy later identified the donor as a red supergiant (RSG; \citealt{Heida2019}). Later, \citet{C18} confirmed that the compact object in SN~2010da is a neutron star by discovering X-ray pulsations. Recent \textit{JWST} observations show the system has returned to its pre-outburst, dust-enshrouded state \citep{Beasor2026}.

Therefore, although the X-ray luminosity of SN~2010da is comparable to that of AT~2016blu, SN~2010da no longer falls within the class of LBV-related SN impostors. This distinction is important: while theoretical models predict SNe from RSG progenitors, a terminal explosion from an LBV is inconsistent with current theoretical expectations \citep{M13}. However, as noted above, SN~2010da is classified as a ULX, and the X-ray luminosity of AT~2016blu lies very close to the ULX threshold. Although this boundary is not strict, it is intriguing that SN impostors and the ULX population, both of which remain poorly understood, appear to be increasingly connected as they are studied in greater detail.

\subsection{Accretion Rate}

Considering the X-ray luminosity is driven by accretion onto a compact object, we can estimate an accretion rate with standard formalism $L=\eta \dot{M}c^2$ \citep{D73}.  Using the X-ray luminosity and assuming $\eta=0.1$ \citep{O12}, we find ${\dot{M}_{\rm acc}} \approx 8\times 10^{-8}~ \mathrm{M_{\odot}\,yr^{-1}}$. However, this assumes that almost all of the accretion luminosity emerges in the X-rays. AT~2016blu's outbursts have a typical brightness of $m_R \approx 17$~mag, corresponding to an absolute magnitude of approximately $-12.8$ after distance and reddening corrections, or an $R$-band luminosity of $\sim10^{7}~\mathrm{L_{\odot}}$, roughly two orders of magnitude larger than the X-ray luminosity ($\sim10^{5}~\mathrm{L_{\odot}}$). If the optical outbursts are themselves powered by accretion, then $\sim$99\% of the accretion luminosity is absorbed and reprocessed into the optical, with only $\sim$1\% escaping as X-rays. The X-ray-derived rate would then underestimate the true accretion rate by a factor of $\sim$100, implying $\dot{M}_{\rm acc} \approx 8\times10^{-6}~\mathrm{M_{\odot}~yr^{-1}}$.  An LBV wind, with a mass-loss rate of $\sim10^{-5}$--$10^{-4}~\mathrm{M_{\odot}~yr^{-1}}$ characteristic of S~Doradus-type variability, could supply the mass required to sustain this accretion rate. Such an accretion rate is more than two orders of magnitude above the Eddington accretion rate for a $1.4~\mathrm{M_{\odot}}$ neutron star, which could be interpreted as favoring a black-hole accretor. However, neutron-star accretors have been observed to reach $\sim 1000$ times the Eddington limit in extreme cases \citep{I17}.

Independent of the accretion rate, we can constrain the wind mass-loss rate using Equations 4–11 of \citet{Lau2016}, adopting $M_* =~33 M_\odot$, $P_{\rm orb} = 113$ days \citep{A25}, $e = 0.4$ \citep{A2316blu}, and assuming the companion is a neutron star with a mass of $M_{\rm NS} = 1.4~M_\odot$. We also adopt wind velocity of $v_w = 500~\mathrm{km~s^{-1}}$, approximately half of the measured H$\alpha$ full width at half-maximum intensity (FWHM $\approx 800$–$1400~\mathrm{km~s^{-1}}$, measured using the procedure of  \citealt{A25}), since the H$\alpha$ FWHM in AT~2016blu is likely dominated by electron scattering rather than bulk wind outflow, as indicated by its persistent Lorentzian profile. With these assumptions, we infer a mass-loss rate of $\dot{M} \approx 3\times10^{-3}~M_\odot~\mathrm{yr^{-1}}$, though we emphasize that this value is highly sensitive to the adopted wind velocity, orbital eccentricity, and companion mass, all of which remain poorly constrained for AT~2016blu. This wind mass-loss rate exceeds even the boosted accretion rate inferred above, indicating that the LBV wind can supply the accreting material. For comparison, \citet{Villar2016} estimated a mass-loss rate of 
$(4\text{--}5)\times10^{-7}~\mathrm{M_{\odot}~yr^{-1}}$ for SN~2010da 
using the inner dust shell radius from their \textsc{dusty} models, while 
\citet{Lau2016} estimated $\dot{M}_w \approx 1.4\times10^{-5}~\mathrm{M_{\odot}~yr^{-1}}$ 
for SN~2010da from its mid-infrared dust emission. Additionally, \citet{Ofek13} estimated a pre-explosion 
mass-loss rate of $\sim10^{-3}$--$10^{-2}~M_\odot~\mathrm{yr^{-1}}$ for 
SN~2009ip using X-ray, optical, and radio observations of shock-CSM interaction following 
its terminal explosion.

\section{Summary} \label{sec:summary}
We present the first X-ray detection of an LBV-related SN impostor, AT~2016blu. The source experienced its first recorded outburst in 2012 \citep{K12} and has since exhibited multiple quasiperiodic outbursts with a period of $\sim 113$ days \citep{A2316blu, A25}. These outbursts are most likely triggered by periastron encounters in an eccentric binary system in which the primary is an LBV, analogous to periastron encounters observed in $\eta$ Carinae and pre-SN outbursts of SN~2009ip.

AT~2016blu was serendipitously observed by \chandra\ in 2001–2002, prior to the first documented outburst, with no X-ray detection. We subsequently monitored the source using KAIT, GSP telescopes, and AAVSO observations, close to the predicted outburst time based on the estimated period. Upon reaching a peak magnitude of 16.6 in March 2026, we triggered \chandra\ ToO observations. \chandra\ observed AT~2016blu across five epochs, with a total exposure time of $\sim$ 100 ks. The source was detected in each epoch, exhibiting a consistent X-ray luminosity. From stacked imaging and power-law spectral modeling, we derive $\log L_{\rm X} \approx 38.64 \pm 0.11$~erg~s$^{-1}$, consistent with accretion onto a compact object.
Given that the primary is a blue straggler with a mass of at least $33~M_\odot$ \citep{A2316blu}, AT~2016blu is an HMXB. Additionally, its luminosity exceeds the Eddington limit for a neutron star and may reach the ultraluminous X-ray regime at peak, making it also a ULX candidate.

Using the X-ray luminosity, we estimate an accretion rate of $\dot{M}_{\rm acc} \gtrsim 8\times10^{-8}~\mathrm{M_\odot~yr^{-1}}$. This is a lower limit, however, because the peak optical luminosity of the outbursts ($\sim10^{7}~\mathrm{L_\odot}$) exceeds the X-ray luminosity by roughly two orders of magnitude, so if the optical outbursts are accretion-powered, the true accretion rate can be $\dot{M}_{\rm acc} \approx 8\times10^{-6}~\mathrm{M_\odot~yr^{-1}}$. Such a rate would be more than two orders of magnitude above the Eddington accretion rate for a $1.4~\mathrm{M_{\odot}}$ neutron star. 

AT~2016blu represents the first known LBV system with a compact companion. Previously studied LBVs either lack X-ray detections or show luminosities consistent with wind–wind collisions in massive-star binaries. Among SN impostors, SN~2010da \citep{Immler2010, Binder2011, Binder2016, Villar2016, Lau2016, Heida2019} also exhibits X-ray emission consistent with accretion and is classified as a ULX, suggesting a possible link between SN impostors and ULXs, both poorly understood populations. However, SN~2010da consists of a RSG, which is expected to end its evolution in an SN explosion. In contrast, LBVs, such as the progenitor of SN~2009ip, may undergo terminal explosions, in tension with single-star evolution models. The discovery of an LBV--compact-object binary undergoing accretion-powered outbursts also connects AT~2016blu to a growing class of models in which massive star--compact object binaries explain a variety of pre-SN transient phenomena \citep{T24}. Expanding the sample of such systems is therefore essential to better understand the evolutionary pathways and physical mechanisms connecting SN impostors, LBVs, and interacting SNe.


\begin{acknowledgments}
This paper employs a list of \chandra datasets, obtained by the {\it Chandra X-ray Observatory}, contained in the \chandra Data Collection (CDC) 591~\dataset[doi:10.25574/cdc.591]{https://doi.org/10.25574/cdc.591}. This research has made use of the NASA/IPAC Infrared Science Archive, which is funded by the National Aeronautics and Space Administration (NASA) and operated by the California Institute of Technology. It also used data from the Asteroid Terrestrial-impact Last Alert System (ATLAS) project, which is funded primarily to search for near-Earth objects (NEOs) through NASA grants NN12AR55G, 80NSSC18K0284, and 80NSSC18K1575; byproducts of the NEO search include images and catalogs from the survey area. The ATLAS science products have been made possible through the contributions of the University of Hawaii Institute for Astronomy, the Queen’s University Belfast, the Space Telescope Science Institute (STScI), the South African Astronomical Observatory, and The Millennium Institute of Astrophysics (MAS), Chile. 

KAIT and its ongoing operation were made possible by donations from Sun Microsystems, Inc., the Hewlett-Packard Company, AutoScope Corporation, Lick Observatory, the U.S.  National Science Foundation (NSF), the University of California, the Sylvia \& Jim Katzman Foundation, and the TABASGO Foundation.
A major upgrade of the Kast spectrograph on the Shane 3~m telescope at Lick Observatory, led by Brad Holden, was made possible through gifts from the Heising-Simons Foundation, 
William and Marina Kast, and the University of California Observatories. Research at Lick Observatory is partially supported by a gift from Google. We appreciate the expert assistance of the staff at
the various observatories where data were obtained.

This work makes use of observations from the Las Cumbres Observatory network. The LCO team is supported by NSF grants AST-2308113 and AST-1911151.
Support for M.A. was provided by the VITA-Origins Fellowship, including funding from the Virginia Institute of Theoretical Astronomy (VITA), supported by the College and Graduate School of Arts and Sciences at the University of Virginia.
A.V.F.’s research group at U.C. Berkeley acknowledges financial   assistance from the Christopher R. Redlich Fund, Gary and Cynthia
Bengier, Clark and Sharon Winslow, Alan Eustace and Kathy Kwan (W.Z. is a Bengier-Winslow-Eustace Specialist in Astronomy), 
Timothy and Melissa Draper, Briggs and Kathleen Wood, Ellyn and Alan Seelenfreund (T.G.B. is Draper-Wood-Seelenfreund Specialist in Astronomy), and numerous other donors.

\end{acknowledgments}

\bibliography{MA}{}
\DeclareRobustCommand{\VAN}[3]{#3}
\bibliographystyle{aasjournal}



\end{document}